\documentclass[sigconf]{acmart}

\usepackage{amsmath,amssymb,amsfonts}
\usepackage{algorithm}
\usepackage{algorithmic}
\usepackage{graphicx}
\usepackage{xspace}
\usepackage{setspace}
\usepackage{xcolor}
\usepackage{tikz}
\usepackage[font=small]{caption} 
\usepackage{hyperref} 
\usepackage{makecell}

\newcommand{\Design}{Stratum\xspace}
\newcommand{\mixtralsmall}{Mixtral 8$\times$7B\xspace}

\newcommand{\dramfull}{Monolithic 3D-Stackable DRAM\xspace}
\newcommand{\dramshort}{Mono3D DRAM\xspace}
\newcommand{\callout}[1]{%
  \tikz[baseline=(char.base)]{
    \node[shape=circle, draw=black, fill=black, text=white, inner sep=0.5pt] (char) {\scriptsize\sffamily\bfseries #1};
  }%
}

\AtBeginDocument{%
  }

\copyrightyear{2025}
\acmYear{2025}
\setcopyright{cc}
\setcctype{by}
\acmConference[MICRO '25]{58th IEEE/ACM International Symposium on Microarchitecture}{October 18--22, 2025}{Seoul, Republic of Korea}
\acmBooktitle{58th IEEE/ACM International Symposium on Microarchitecture (MICRO '25), October 18--22, 2025, Seoul, Republic of Korea}\acmDOI{10.1145/3725843.3756043}
\acmISBN{979-8-4007-1573-0/2025/10}

\begin{document}

    



\author[Y. Pan]{Yue Pan}
\authornote{Equal contribution}

\affiliation{
  \institution{University of California, San Diego}
  \city{La Jolla}
  \country{United States}
}
\email{yup014@ucsd.edu}

\author[Z. Xia]{Zihan Xia}
\authornotemark[1]
\affiliation{
  \institution{University of California, San Diego}
  \city{La Jolla}
  \country{United States}
}
\email{z5xia@ucsd.edu}

\author[P. Hsu]{Po-Kai Hsu}
\affiliation{
  \institution{Georgia Tech}
  \city{Atlanta}
  \country{United States}
}
\email{pokai.hsu@gatech.edu}

\author[L. Hu]{Lanxiang Hu}
\affiliation{
  \institution{University of California, San Diego}
  \city{La Jolla}
  \country{United States}
}
\email{lah003@ucsd.edu}

\author[H. Kim]{Hyungyo Kim}
\affiliation{
  \institution{\mbox{University of Illinois, Urbana-Champaign}}
  \city{Urbana}
  \country{United States}
}
\email{hyungyo2@illinois.edu}

\author[J. Sharda]{Janak Sharda}
\affiliation{
  \institution{Georgia Tech}
  \city{Atlanta}
  \country{United States}
}
\email{jsharda3@gatech.edu}

\author[M. Zhou]{Minxuan Zhou}
\affiliation{
  \institution{Illinois Institute of Technology}
  \city{Chicago}
  \country{United States}
}
\email{mzhou26@illinoistech.edu}

\author[N. S. Kim]{Nam Sung Kim}
\affiliation{
  \institution{\mbox{University of Illinois, Urbana-Champaign}}
  \city{Urbana}
  \country{United States}
}
\email{nskim@illinois.edu}

\author[S. Yu]{Shimeng Yu}
\affiliation{
  \institution{Georgia Tech}
  \city{Atlanta}
  \country{United States}
}
\email{shimeng.yu@ece.gatech.edu}

\author[T. Rosing]{Tajana Rosing}
\affiliation{
  \institution{University of California, San Diego}
  \city{La Jolla}
  \country{United States}
}
\email{tajana@ucsd.edu}

\author[M. Kang]{Mingu Kang}
\affiliation{
  \institution{University of California, San Diego}
  \city{La Jolla}
  \country{United States}
}
\email{mingu@ucsd.edu}


\begin{abstract}

As Large Language Models (LLMs) continue to evolve, Mixture of Experts (MoE) architecture has emerged as a prevailing design for achieving state-of-the-art performance across a wide range of tasks. MoE models use sparse gating to activate only a handful of expert sub-networks per input, achieving billion-parameter capacity with inference costs akin to much smaller models. However, such models often pose challenges for hardware deployment due to the massive data volume introduced by the MoE layers. 
To address the challenges of serving MoE models, we propose \Design, a system–hardware co-design approach that combines the novel memory technology \dramfull{} (\dramshort{}), near-memory processing (NMP), and GPU acceleration. The logic and \dramshort{} dies are connected through hybrid bonding, whereas the \dramshort{} stack and GPU are interconnected via silicon interposer.
\dramshort offers higher internal bandwidth than HBM thanks to the dense vertical interconnect pitch enabled by its monolithic structure, which supports implementations of higher-performance near-memory processing. Furthermore, we tackle the latency differences introduced by aggressive vertical scaling of \dramshort~along the $z$-dimension by constructing internal memory tiers and assigning data across layers based on access likelihood, guided by topic-based expert usage prediction to boost NMP throughput.
The \Design system achieves up to 8.29$\times$ improvement in decoding throughput and 7.66$\times$ better energy efficiency across various benchmarks compared to GPU baselines.

\end{abstract}

\title{Stratum: System-Hardware Co-Design with Tiered \dramfull for Efficient MoE Serving}



\maketitle

\section{Introduction}

Transformer-based Large Language Models (LLMs) have become central to a wide range of applications, delivering state-of-the-art performances across diverse domains~\cite{attention_is_all_you_need, llama3, vit, zhang2022opt, xai_grok3, mixtral, gpt4_report, yang2024qwen2, diffusion, deepseek-v3, deepseek-r1}.
To improve various task performances, LLMs are reaching unprecedented scales, with models such as LLaMA 3.1 (405B)~\cite{llama3}, DeepSeek-V3 (671B)~\cite{deepseek-v3}, and Kimi-K2 (1T)~\cite{kimi-k2} pushing the boundaries of model size and performance.
Training and deploying these large models present significant challenges to the underlying infrastructure, particularly in terms of memory capacity and compute capability.

\begin{figure}
    \centering
    \includegraphics[width=\columnwidth]{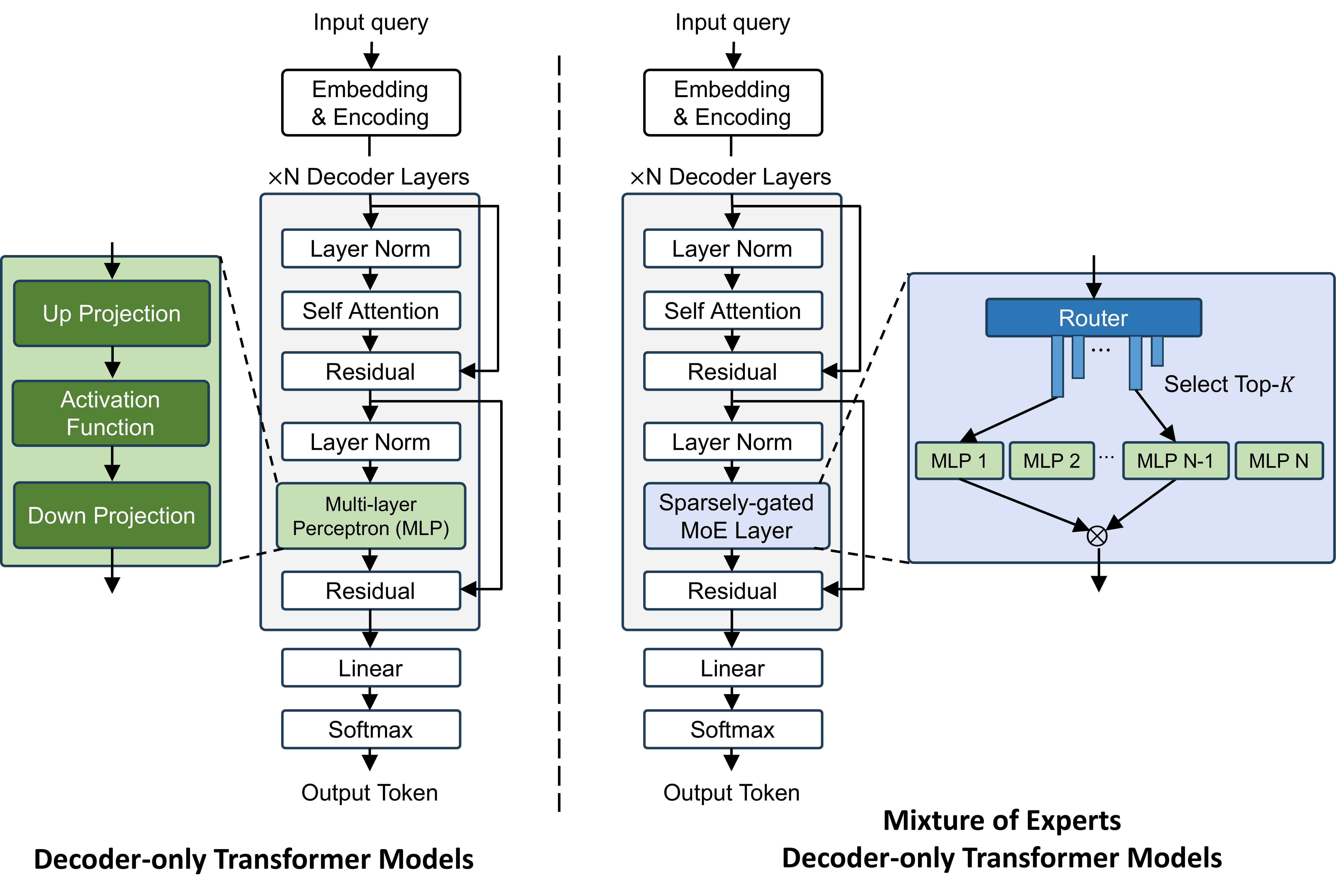}
    \caption{Architectures of dense transformer-based LLM (left) and Mixture of Experts (MoE) LLM (right).}
    \label{fig:moellm}
\end{figure}

Among various efforts to reduce the inference cost, 
exploiting activation sparsity offers a promising solution by directly reducing the computational and data movement demands. One of the most widely adopted approaches is the Mixture of Experts (MoE) architecture~\cite{deepseek-v3, gpt4_report, olmoe, mixtral, glam, dbrx2024, xai_grok3, switch-transformer, llama4}, which replaces conventional dense Multi-Layer Perceptron (MLP) blocks with a pool of expert MLPs that are sparsely selected during inference, as illustrated in Figure~\ref{fig:moellm}. MoE models utilize a routing mechanism to activate only a small subset of experts per token during inference. Since MLP dominates the overall model size, this selective activation leads to substantial savings in both inference and training costs~\cite{moe-scaling-law}. 
As a result, the MoE architecture has become a preferred choice in many state-of-the-art LLMs.

While MoE models reduce practical memory access and computation requirements, they do not address the overall size of the model. The rapid growth in model size necessitates high-bandwidth and high-density memory technologies. Along this line, die-stacked High Bandwidth Memory (HBM) has emerged as the dominant solution in high-performance GPUs such as the NVIDIA A100 and H100~\cite{a100, h100}, achieving high density per footprint with six stacked DRAM dies and 1024-bit I/O interfaces, delivering up to 800~GB/s of memory bandwidth per stack to the GPU compute die via silicon interposers. 
Although HBM offers increased bandwidth compared to conventional 2D DRAMs, the bandwidth available through the interposer remains insufficient. This limitation often leads to underutilization of GPU computing resources, particularly for memory-bound operations such as LLM decoding ~\cite{attacc_asplos24}. To mitigate the memory wall between HBM and the GPU, recent approaches have adopted near-memory processing (NMP) for LLM inference~\cite{he2020newton, zhou2022transpim, hbm-pim, attacc_asplos24, duplex_micro24, heo2024neupims, near-hbm-cnn}. Prior studies~\cite{heo2024neupims, attacc_asplos24, zhou2022transpim, duplex_micro24} have utilized NMP units to compute attention during the decoding stage by placing the computing logic on the HBM base die. However, the NMP on the base die still suffers from limited bandwidth due to vertical data traversal through a constrained number of TSV I/O connections. To mitigate this limitation, prior work has integrated compute units directly into the memory dies to exploit extensive internal memory bandwidth~\cite{hbm-pim, zhou2022transpim, drisa, attacc_asplos24, near-hbm-cnn, primate}, commonly known as processing in memory (PIM). However, compute logic embedded in DRAM dies suffers from expensive intra-memory data transmission and large performance-area-power (PPA) overhead of implementing logic using the DRAM technology, as DRAM dies are inherently optimized for storage rather than computation~\cite{drisa}. Moreover, integrating logic and memory on the same die introduces additional thermal concerns and manufacturing overheads.

As a strong alternative to HBM, \dramfull, referred to as \dramshort throughout this paper, has recently emerged as a promising solution for continued DRAM scaling beyond sub-10-nanometer technologies. It offers improved vertical integration through a cost-effective fabrication process that eliminates costly TSV and bonding processes, gaining growing attention in both industry and academia~\cite{3DDRAM_Samsung, 3DDRAM_SK, 3DDRAM_Stanford,3DDRAM_GT}. By fabricating multiple additional DRAM layers sequentially on the same wafer, \dramshort achieves higher density without a proportional increase in cost per bit, making it an attractive candidate for future high-capacity memory systems. 
Compared to HBM-based NMP, \dramshort-based NMP introduces key architectural benefits. \dramshort offers significantly greater internal bandwidth due to its monolithic construction within DRAM and direct face-to-face hybrid bonding between DRAM and logic dies, leveraging the full chip area. On the other hand, TSVs in HBM require a certain area on both the logic base die and DRAM dies as vertical interconnects. The TSV area cannot be unbounded, thus limiting the HBM internal bandwidth. Moreover, hybrid bonding pitch of 1~$\mu m$~\cite{CuCuHyBond} has around 5$\times$ finer pitch for vertical interconnects than HBM~\cite{HB_isca24}, offering denser internal connectivity. The higher internal bandwidth of \dramshort can enable stronger NMP capability with the logic-die implementation than prior HBM-based memory-die NMP architectures.
In addition, thinner dies and improved vertical thermal conduction enabled by monolithic integration enhance heat dissipation, supporting higher power density and allowing a larger power budget for NMP. 

Despite the numerous potential benefits offered by \dramshort, fully leveraging its advantages presents several critical challenges. Recent studies have demonstrated the feasibility of integrating several hundred vertically stacked layers through sequential layer fabrication~\cite{3DDRAM_Stanford, 3DDRAM_GT}. However, such aggressive vertical scaling inherently leads to substantial variability in access latencies across different layers. Adopting a simplistic design based on the worst-case latency significantly undermines the available internal bandwidth. Additionally, the drastically increased density of vertical interconnects, enabled by the fine-pitch monolithic 3D integration, facilitates simultaneous access to large volumes of data. Consequently, a carefully tailored data mapping strategy is essential to effectively harness local \dramshort~bank bandwidth while minimizing inter-bank and inter-channel data access. Furthermore, given the extremely high local DRAM data access bandwidth, the overhead of on-chip communication between processing units can become comparable to the computation latency if data is mapped inefficiently. Therefore, achieving a balanced overlap between computation and communication is crucial for minimizing the overall execution time. 

To address the challenges in serving large MoE models, we propose the \Design system that integrates \dramshort, NMP, and GPU. This work makes the following key contributions:

\noindent$\bullet$  For the first time, we propose a system-hardware co-design solution \Design~for MoE serving that leverages \dramfull. Our approach heterogeneously integrates high-density \dramshort~dies with high-performance logic dies via 3D hybrid bonding, and further integrates this \dramshort~stack with GPUs using a 2.5D silicon interposer. This architecture serves as a high-throughput and cost-effective alternative to conventional GPU-HBM-based MoE serving systems. 

\noindent$\bullet$  At the hardware level, we introduce an in-memory tiering mechanism that exploits the inherent access latency variations across \dramshort~layers resulting from vertical scaling. Additionally, we propose an NMP processor tailored for hybrid-bonding-based \dramshort, incorporating optimized data mapping and communication strategies for both expert and attention execution.

\noindent$\bullet$  At the system level, we observe the nonuniform activation frequency of experts depending on user request topics. Based on this, we classify experts into hot and cold categories and assign them to fast and slow tiers of \dramshort, respectively. The proposed topic-aware serving system queues and dispatches requests according to their topics, predicted by our and lightweight topic classifier, while adhering to defined service-level objectives (SLOs).

\noindent$\bullet$  Cross-layer evaluations (device, circuit, algorithm, and system) demonstrate that \Design achieves up to 8.29$\times$ better decoding throughput and 7.66$\times$ better energy efficiency in practical MoE serving scenarios, compared to state-of-the-art GPU-baselines.

\section{Background}

\subsection{Monolithic 3D-Stackable DRAM}\label{sec:background 3d dram}

\dramshort is a promising technology for continued DRAM scaling, drawing significant attention from both academia and industry~\cite{hbm-pim, zhou2022transpim, drisa, attacc_asplos24, near-hbm-cnn}. 
Compared to conventional 2D DRAM technologies, it offers significantly higher memory density by leveraging vertical scaling—enabled by advanced techniques such as nanosheet field-effect transistors (FETs), which provide tighter gate control and support stacked channel architectures, and fabrication techniques inspired by 3D NAND Flash processes, including layer-by-layer deposition, high-aspect-ratio etching for ultra-thin dielectric isolation, and dense vertical integration~\cite{3DDRAM_Samsung,3DDRAM_Stanford,3DDRAM_SK, 3DDRAM_GT}.

\begin{figure}
    \centering
    \includegraphics[width=\columnwidth]{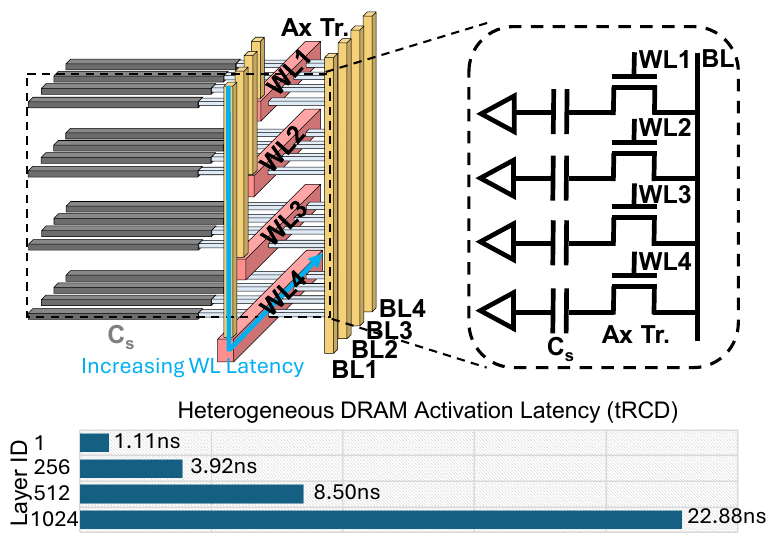}
    \caption{\dramfull with vertically stacked horizontal 1T1C DRAM cells. Bitlines are vertically routed to avoid sense margin variations, and wordlines are routed through staircases. The activation latency varies by layers due to wordline staircases.}
    \label{fig:3ddram_tech}
\end{figure}

\dramshort employs monolithic 3D stackable horizontal 1T1C DRAM cells, incorporating wordline (WL) staircases and vertically connected bitlines (BL) to interconnect memory cells across multiple layers, as seen in Figure~\ref{fig:3ddram_tech}. While HBM incurs high costs due to low manufacturing yield from TSV fabrication and the sophisticated packaging required for die stacking, \dramshort offers cost advantages through improved scalability by avoiding TSVs and leveraging monolithic 3D integration, which sequentially constructs additional DRAM layers on the same wafer. \dramshort also achieves thermal benefit using thinner dies and improved vertical thermal conduction enabled by monolithic integration.

\begin{figure}
    \centering
    \includegraphics[width=\columnwidth]{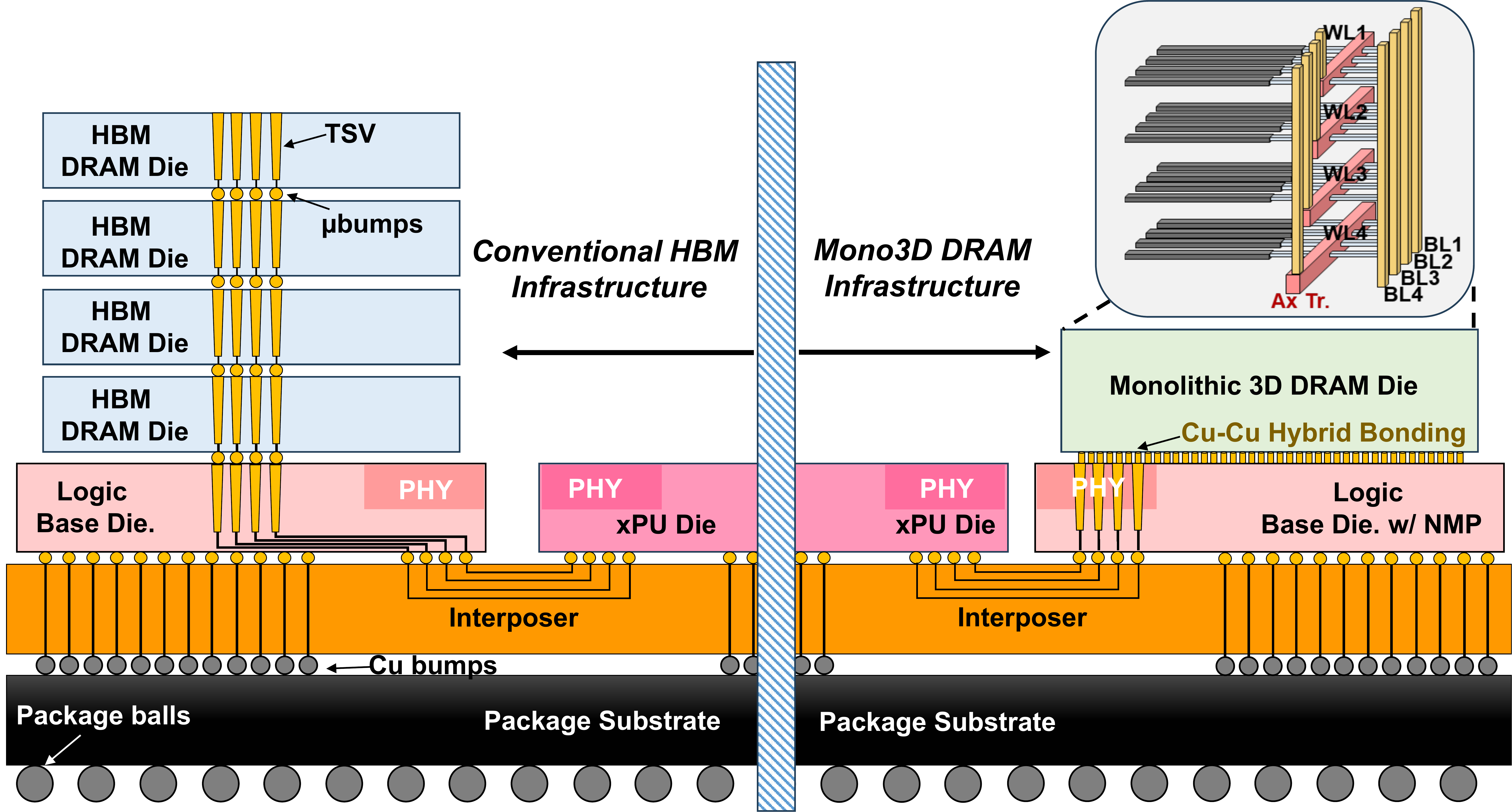}
    \caption{HBM versus \dramshort on 2.5D integration platform with a xPU die. The HBM and \dramshort are attached to the logic base die through TSVs and Cu-Cu hybrid bonding, respectively. 
    }
    \label{fig:hbmvs3ddram}
\end{figure}

On top of its cost and thermal benefits, \dramshort also delivers enhanced memory bandwidth to the logic layer. It leverages heterogeneous integration~\cite{3DDRAM_SK, 3DDRAM_GT} and employs Cu–Cu hybrid bonding for high-speed data transfer between memory cells and logic peripherals. Figure~\ref{fig:hbmvs3ddram} compares \dramshort with HBM on the same 2.5D integration platform. HBM’s internal bandwidth is constrained by TSVs, which have a coarse pitch of 10~$\mu$m~\cite{TSV}, resulting in limited bandwidth and significant area overhead that reduces memory density.
In contrast, \dramshort utilizes Cu–Cu hybrid bonding between DRAM and logic base dies with a much finer pitch of 1~$\mu$m~\cite{CuCuHyBond}, connected via back-end-of-line (BEOL) metal routing, to achieve exceptionally high internal bandwidth.

Despite its higher internal bandwidth, \dramshort, as shown in Figure~\ref{fig:hbmvs3ddram}, still has external bandwidth limitations similar to HBM due to the limited bandwidth of the interposer I/O interface. Additionally, prior work ~\cite{MICRO17_FineGrainedDRAM} highlights the significant energy consumption incurred during data transfers to the external processor, including routing across the logic base die and through the interposer I/O interface. 
These inefficiencies underscore the necessity of NMP integration on the logic die alongside \dramshort to utilize internal bandwidth and improve energy efficiency.

Despite the potential for exceptional memory capacity in \dramshort, its vertical scalability is limited by substantial variation in access latency across layers. As shown in Figure~\ref{fig:3ddram_tech}, WLs at the bottom of the staircase structure experience increased parasitic capacitance and resistance, resulting from the linearly extended WL routing. This latency imbalance becomes significant when \dramshort is scaled to hundreds of layers. Rather than designing around the worst-case access latency, system-level performance can be improved by embracing this latency heterogeneity. This challenge naturally motivates an architectural approach dubbed \textit{in-memory tiering}, discussed in detail in \S\ref{sec:system}. 
Note that the scaling trend of \dramshort aligns with that of 3D NAND Flash, as \dramshort leverages similar fabrication processes that have already been scaled beyond 400 layers~\cite{Samsung_3DNAND_400}. Furthermore, recent white papers suggest the feasibility of extending this scaling to 500 to even 1000 layers~\cite{neo_xdram, lam3DNAND1000L}. Given these advancements and the projected trajectory of vertical scaling, we assume up to 1024 wordline (WL) stacks to reflect the near-future feasibility.

\subsection{Mixture of Expert LLMs}\label{sec: moe}

As indicated by LLM scaling laws~\cite{kaplan2020scalinglawsneurallanguage}, the accuracy of dense transformer models improves with size, but so do their training and serving costs. Recent MoE models, such as OLMoE~\cite{olmoe}, Mixtral~\cite{mixtral}, Deepseek V3~\cite{deepseek-v3}, Time MoE~\cite{shi2024timemoebillionscaletimeseries}, DBRX~\cite{dbrx2024}, LLaMA-4~\cite{llama4}, and Kimi-K2~\cite{kimi-k2}, offer a compelling alternative by activating only a small subset of experts per token. This sparse activation improves training scalability and enables large parameter counts without proportional increases in pre-training cost~\cite{moe-scaling-law}, while keeping inference costs comparable to smaller dense models~\cite{switch-transformer}. On the other hand, MoE models require a routing mechanism, where a gating network computes expert assignment scores from token representations (FFN input or intermediate activations) using learned router parameters that determine sparse expert selection patterns~\cite{switch-transformer}. Each token is then dispatched to its selected expert(s) for independent processing, and when multiple experts are used per token, their outputs are combined—typically via weighted aggregation using the routing scores—to produce the final output of the layer~\cite{mixtral, deepseek-v3, switch-transformer}.

The switching nature of MLP modules in MoE models introduces unique hardware deployment challenges. First, MoE models are large, with expert weights dominating the total size, e.g., over 95\% of the model in \mixtralsmall~\cite{mixtral}, placing substantial pressure on GPU memory. Second, expert usage varies dynamically for each token and is unknown beforehand, leading to load imbalance when experts are distributed across different computing units~\cite{deepseek-v3}. Recent efforts aim to reduce communication overhead by predicting expert usage in advance. ExpertFlow~\cite{he2024expertflow} employs a lightweight surrogate model to forecast routing paths, while MoE Infinity~\cite{moe-infinity} uses cross-layer activation profiling to statistically predict expert selection. In hybrid GPU and near-memory processing systems, Duplex~\cite{duplex_micro24} dynamically dispatches expert computation to either GPU or NMP units based on the latency models and batch size.

During training, MoE models typically include an expert imbalance loss to prevent starvation, where one or more experts are selected far less frequently, thereby encouraging more uniform expert utilization~\cite{switch-transformer, mixtral}. However, as training progresses, domain specialization tends to emerge naturally among experts~\cite{li2023accelerating, child2019generating, yao2024exploitinginterlayerexpertaffinity}. This specialization becomes increasingly pronounced as the number of experts increases and shared experts are introduced, consolidating common knowledge and enhancing the domain specificity of the routed experts~\cite {olmoe, deepseek-v3, dai2024deepseekmoe, llama4}. Building on this observation, recent work has explored leveraging expert affinity to specific domains to accelerate inference in GPU-only environments~\cite{yao2024exploitinginterlayerexpertaffinity, wei2024aptmoe, go2025moetuner}.

\begin{figure}
    \centering
    \includegraphics[width=\columnwidth]{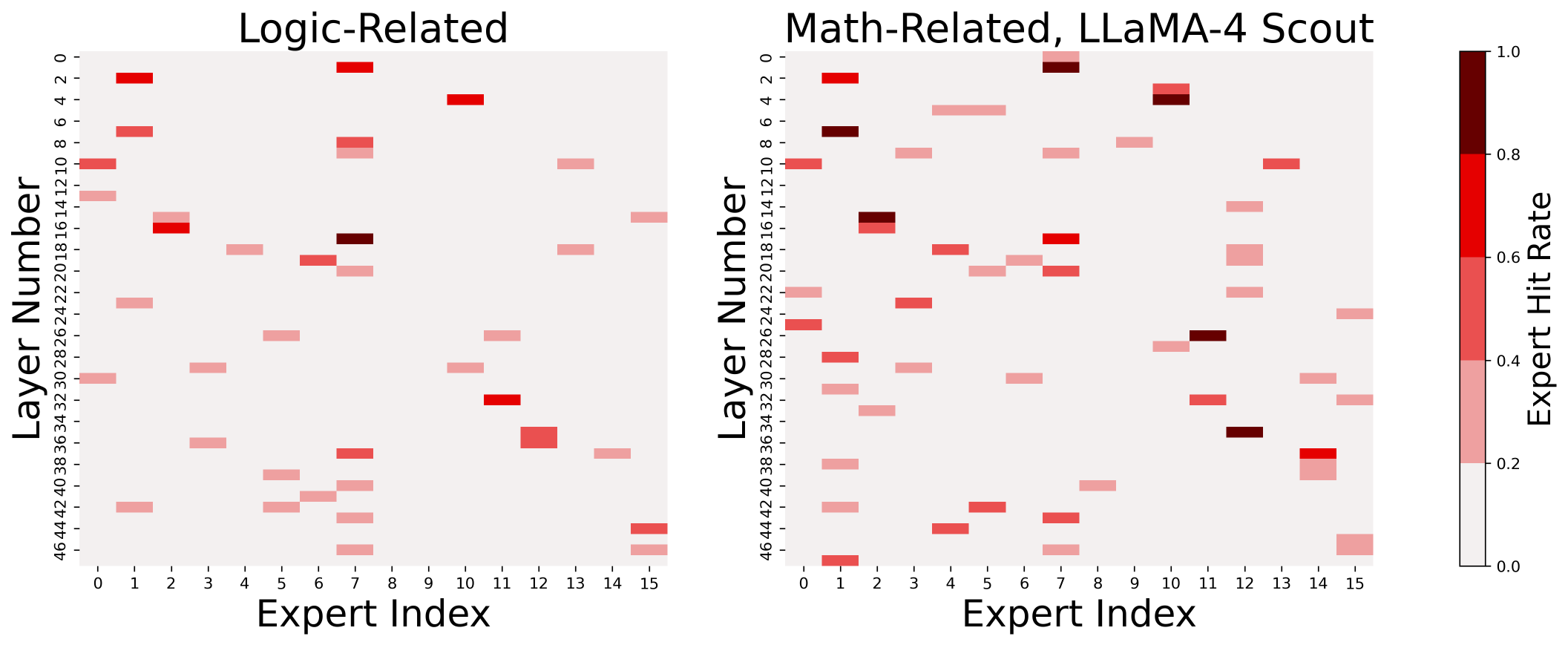}

    \caption{Expert hit profiling from LLaMA-4 Scout (16 Experts).
    }
    \label{fig:expert specialization}
\end{figure}

We profile and observe that the expert usage has a distinct relationship with the topic of the query: a particular topic activates certain experts significantly more frequently. An example is shown in Figure~\ref{fig:expert specialization}, where LLaMA-4 Scout exhibits over 90\% domain-specific expert affinity on math- and logic-related topics within MMLU subsets. In our serving system, we exploit topic-specific expert affinity by first conducting offline profiling to collect statistics on expert hit rates (i.e., usage probabilities) across various topics. During online serving, a lightweight topic classifier in the scheduler assigns topic labels to all incoming queries in a batch. Based on this classification, the system maps frequently used experts to faster Mono3D DRAM layers to optimize access latency, as discussed in \S\ref{sec:alg-sys-codesign}.

\section{\Design Overview}\label{sec:system}

\subsection{System Overview}\label{sec: system overview}

\label{sec: serving}

\begin{figure}
    \centering
    \includegraphics[width=0.8\columnwidth]{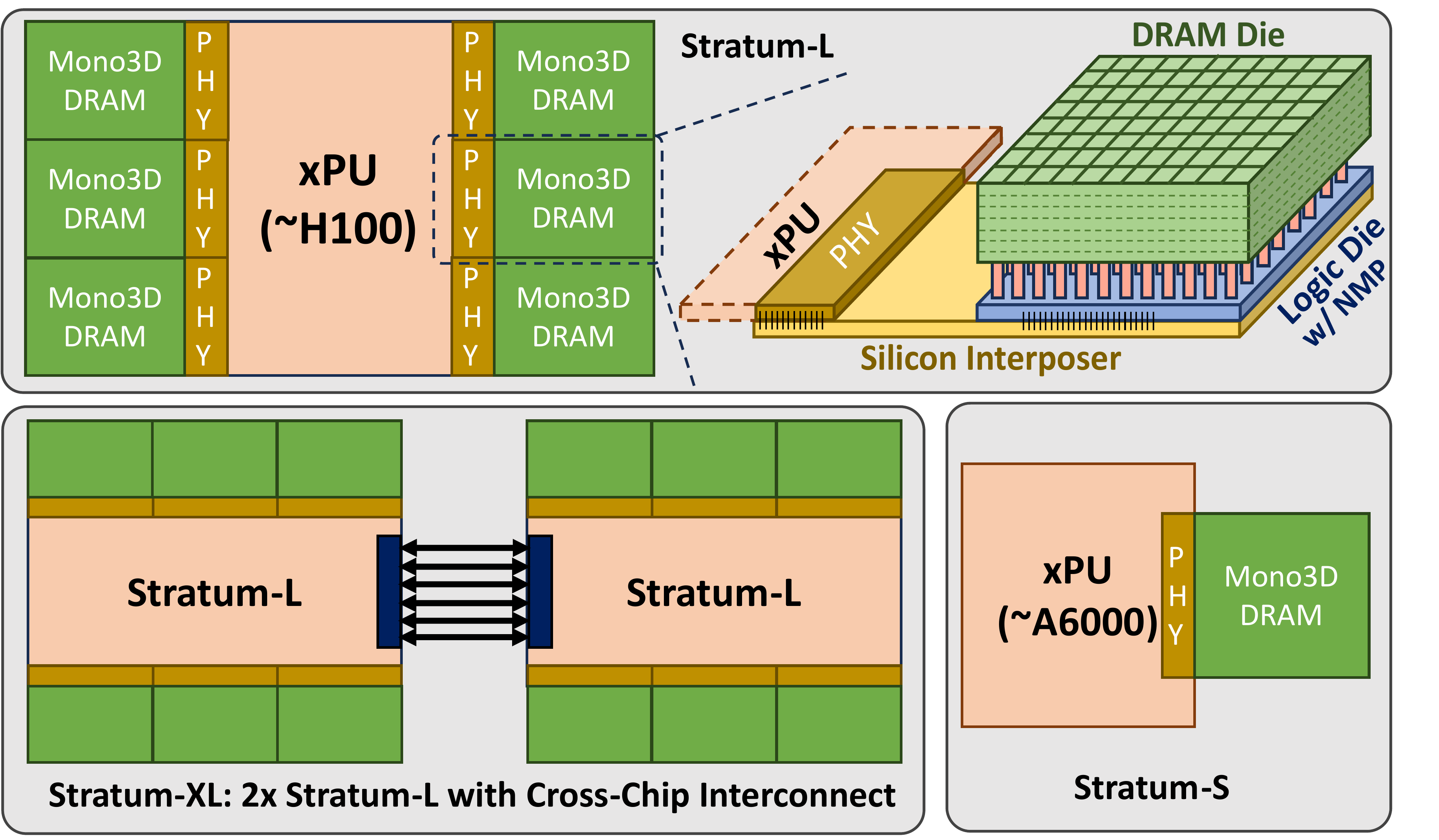}
    \caption{Example \Design configurations.}
    \label{fig:overall_system}
\end{figure}

\begin{figure}
    \centering
    \includegraphics[width=\columnwidth]{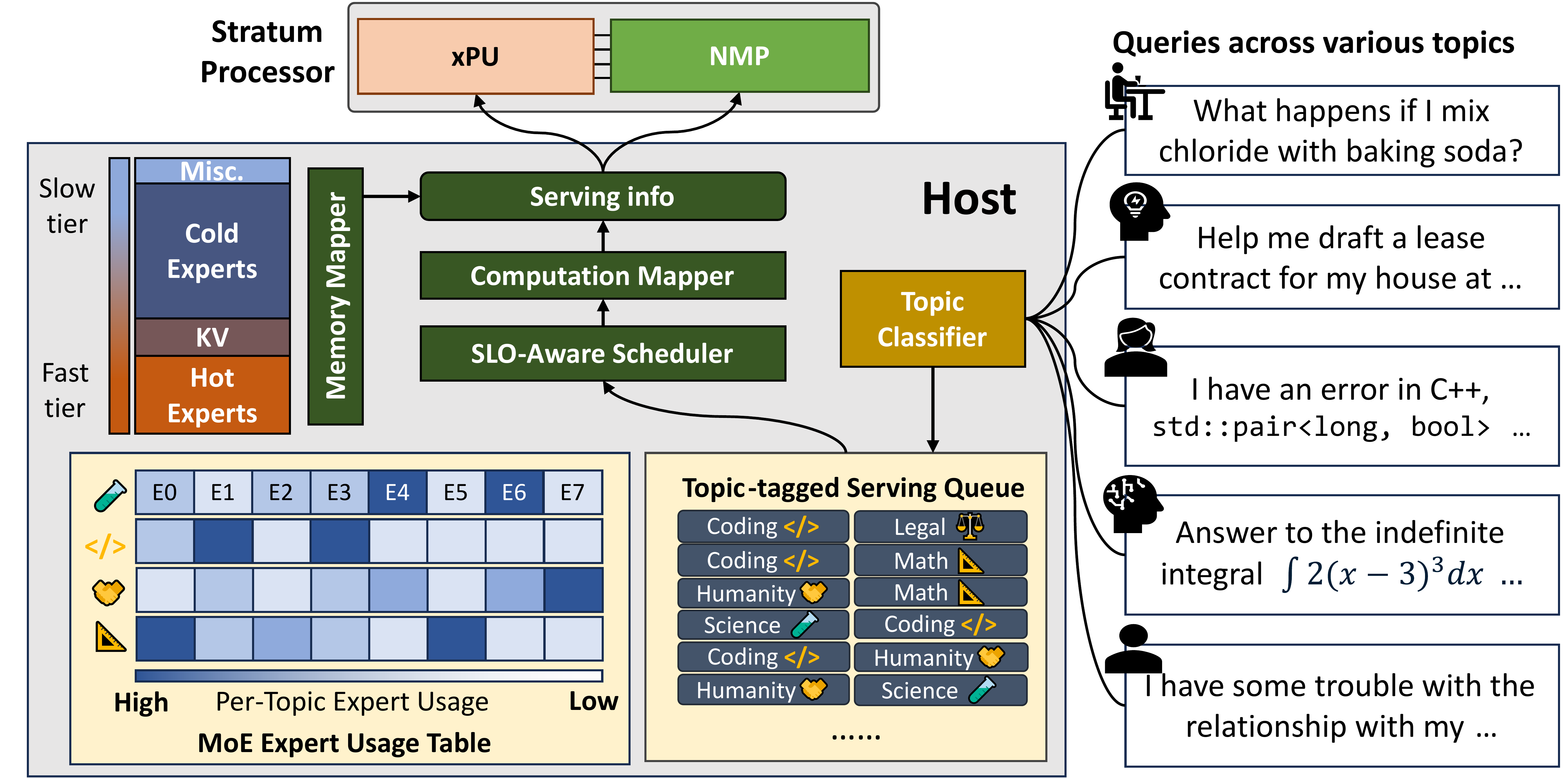}
    \caption{Serving system based on \Design.}
    \label{fig:serving system}
\end{figure}

\begin{figure*}
    \centering
    \includegraphics[width=1.85\columnwidth]{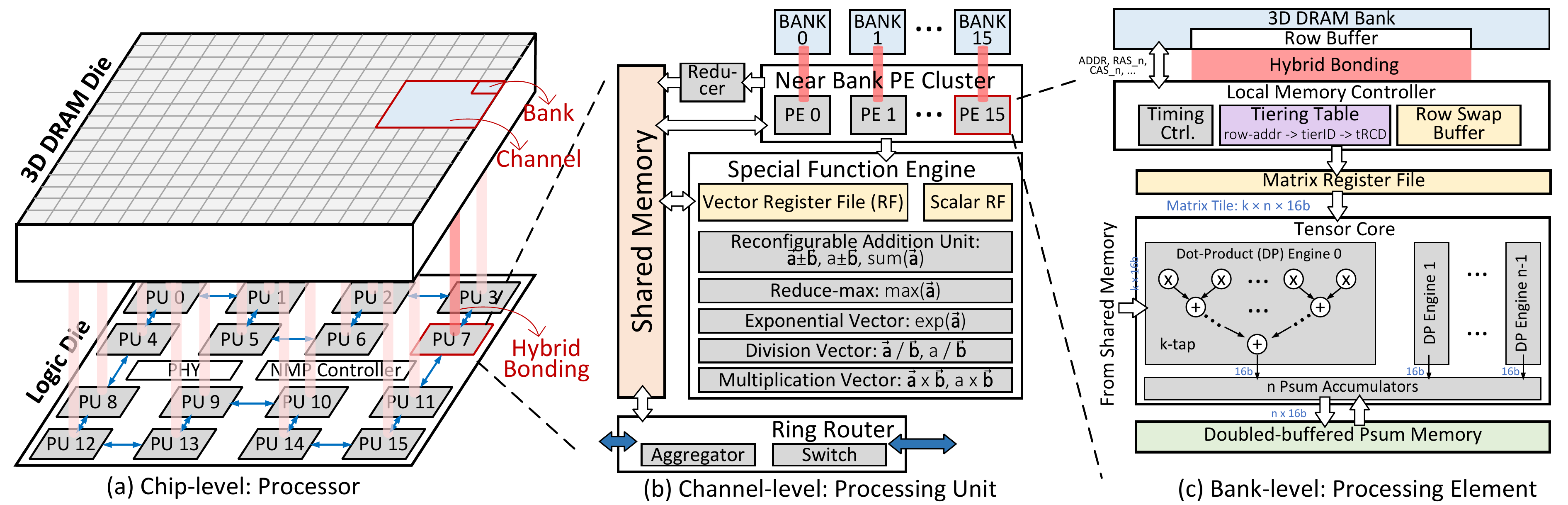}
    
    \caption{\Design~NMP architecture. (a) Overview of the processor at the chip level. Microarchitectures of (b)  the processing unit (PU) at the channel level, and  (c)  the processing element (PE) at the bank level.}
    \label{fig:hardware_arch}
\end{figure*}

The \Design processing system consists of an xPU die and a configurable number of \dramfull chips, interfaced through silicon interposers, with near-memory computing capabilities. 
We demonstrate three different example configurations (Figure~\ref{fig:overall_system}) to accommodate models of varying sizes, using different numbers of \dramshort chips. \textit{Stratum-L} uses an NVIDIA H100 compute die as the xPU die with six \dramshort chips interconnected through interposers. \textit{Stratum-S} uses a NVIDIA RTX A6000 die as the xPU die with a single \dramshort chip providing 32GB memory. \textit{\Design-XL} consists of two \textit{\Design-L} modules, providing a total of 384 GB of memory for serving larger models. These configurations suit diverse compute and memory requirements, and can scale up using cross-chip interconnects like NVLink~\cite{nvlink}.

Each \dramshort chip consists of a memory die on top and a logic die at the bottom, which are interconnected by Cu-Cu hybrid bonding to provide high internal bandwidth. Additionally, to exploit access latency differences across the vertical layers of \dramshort, we introduce internal memory tiering within the memory die. The bottom logic die implements a powerful near-memory processor (NMP) to support LLM inference without always fetching data to the host processor, as detailed in \S\ref{sec: nmp_arch}. 

Figure~\ref{fig:serving system} describes the flow of a serving system based on \Design. In a realistic serving scenario, queries submitted by users are of varying topics. 
When users send inference requests, the host processor uses a lightweight topic classifier to determine the topic of the query. 
These requests are then enqueued in the serving queue with a topic tag. 
Periodically, the scheduler groups inference requests from the serving queue and later dispatches them to the \Design processing system. To enhance user experience, a key Service-Level Objective (SLO) is Time to First Token (TTFT), which ensures that a request does not wait too long before processing begins. When SLO permits, the scheduler prioritizes batching requests of the same topic to maximize the benefits of expert placements. The memory mapper constructs the aggregated expert hit prediction for the batch by consulting the pre-profiled expert usage table and produces a target placement as a mapping between experts to \dramshort layers. Expert swaps are executed before every new batch with different topic tags to meet the target layout. 
Considering the arithmetic intensity of each stage, the Computation Mapper assigns the prefill phase to xPU and the decode phase to the \Design NMP, following a similar strategy as in~\cite{attacc_asplos24}. Additionally, the lightweight topic classification is executed by the host processor.

\subsection{\Design Near Memory Processing}\label{sec: nmp_arch}
Figure~\ref{fig:hardware_arch} illustrates the architecture of \Design NMP, which organizes processing components across multiple levels of the memory hierarchy—including chip, channel, and bank levels—to exploit the benefits of 3D integration. This architectural decision targets the acceleration of attention and expert computations, which are fundamental bottlenecks in MoE models.

Figure~\ref{fig:hardware_arch}(a) illustrates the integration of the logic die processor with the \dramshort die. The logic die consists of multiple processing units (PUs), each coupled with a dedicated \dramshort channel. These PUs interconnect via a bidirectional ring-based on-chip network designed to optimize data communication patterns in LLM workloads, such as reduce-scatter and all-gather. Note that the ring network is only utilized in NMP mode. In regular memory operation mode, the logic die NMP remains inactive, ensuring minimal interference with traditional memory access patterns. In NMP mode, the xPU streams inputs (e.g., queries, hidden token vectors, etc.) to reserved rows in \dramshort~banks with a standard DRAM interface. Upon computation completion, the xPU retrieves processed results by accessing the dedicated address space. 

Each PU aims to handle data assigned to its respective DRAM channel to avoid cross-channel DRAM access—a critical consideration given the massive volume of vertical routing between \dramshort and the logic die. Figure~\ref{fig:hardware_arch}(b) presents the PU microarchitecture, consisting of a near-bank processing element (PE) cluster, a shared memory, a special function engine, a ring router, and a reducer. The near-bank PE cluster integrates multiple PEs optimized for both GeMM and GeMV operations. The intra-channel reducer implemented with parallel reduction trees aggregates partial sums (psums) across multiple PEs within the channel as required. The ring router incorporates a local switch for efficient data routing during inter-PU communication and an aggregator for in-situ data reduction. Incoming data streams can be immediately accumulated in the router without going through the shared memory. The accumulated results can be stored locally in the PU  or forwarded to neighboring PUs as needed. The special function engine performs special operations such as \texttt{Softmax} for attention mechanisms and other common activation functions (e.g., \texttt{SiLU}, \texttt{GeLU}) in expert layers. It includes a vector register file, a scalar register file, and multiple arithmetic units. Operating in a single-instruction-multiple-data (SIMD) manner, the special function engine maximizes data reuse by decomposing complex functions into simple primitives and sourcing and storing operands or intermediate results within the vector and scalar register files.

At the bank level, detailed in Figure~\ref{fig:hardware_arch}(c), each PE is designed to execute GeMM and GeMV operations. The bank-level PE consists of a tensor core integrated with specialized memory components: a matrix register file, a psum memory, and a simple local memory controller. The memory controller, directly interfacing with its corresponding DRAM bank, dynamically translates row addresses to specific memory tier identifiers through a programmable tiering table, enabling adaptive DRAM latency control (\texttt{tRCD}) for performance optimization. The row swap buffer stores temporary row data to support tier-to-tier data movement without requiring explicit external data fetching. The tensor core incorporates $n$ parallel $k$-tap dot-product engines and $n$ local accumulators. The double-buffered psum memory structure concurrently supports intermediate result accumulation and output transfers. The processed outputs can be delivered to the special function engine for element-wise function evaluation or returned to the channel-level shared memory for subsequent computational steps.

Stratum's architecture, specifically optimized for hybrid bonding-based \dramshort integration, differs from HBM-centric NMP approaches such as AttAcc~\cite{attacc_asplos24}, Neupims~\cite{heo2024neupims}, and Duplex~\cite{duplex_micro24}. The on-chip ring network is designed to support MoE inference communication patterns (e.g., all-gather, reduce-scatter), eliminating the centralized global buffer and crossbar used in Duplex~\cite{duplex_micro24}, which improves scalability and simplifies physical design. Unlike Duplex~\cite{duplex_micro24} and AttAcc~\cite{attacc_asplos24}, which rely on dedicated \texttt{Softmax} units, our SIMD-based engine executes general non-linear operators with programming instructions. In addition, the processor is fully implemented on the logic die and hybrid-bonded to the \dramshort die, avoiding the DRAM fabrication process constraints and TSV bandwidth limitations observed in AttAcc~\cite{attacc_asplos24} and Neupims~\cite{heo2024neupims}. At the circuit level, \Design introduces \dramshort-specific primitives—including tiering tables and row swap buffers—to exploit tiered memory latency and accelerate expert migration for MoE model serving.

\section{\Design Operator Mapping and Execution}
\label{sec: operator_map_exe}

\subsection{Expert Processing}\label{sec:expert processing}
\begin{figure}
    \centering
    \includegraphics[width=\columnwidth]{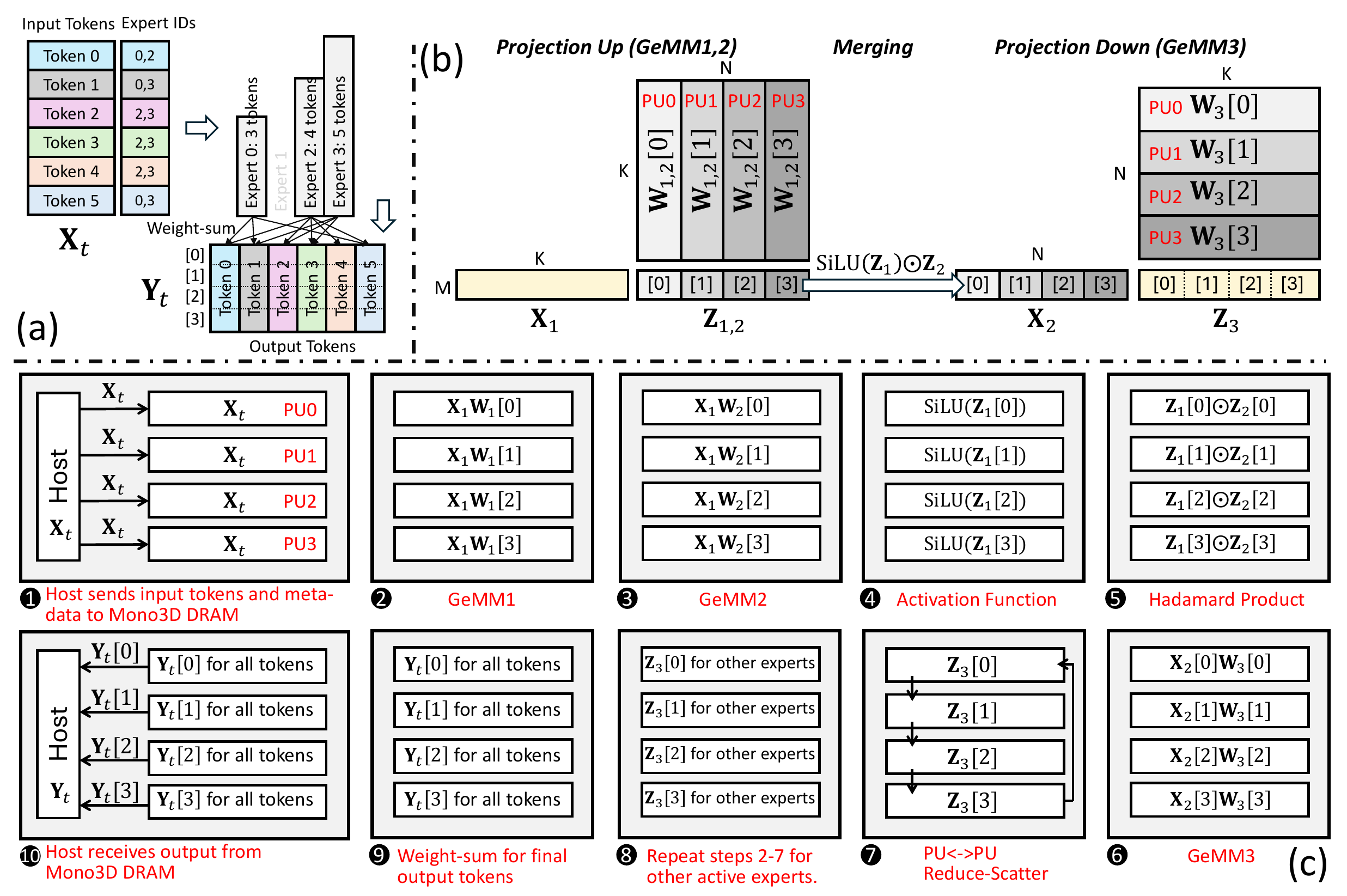}
    \caption{(a) Example of MoE's token-to-expert mapping. (b) The computation stages of an expert with $M$ routed tokens and matrix partition, assuming four PUs for simplicity. (c) The step-by-step execution of the MoE layer in \Design.}
    \label{fig:expert_execution}
\end{figure}

The execution flow of an MoE layer consists of three main stages: token routing, expert computation, and result aggregation. As illustrated in Figure~\ref{fig:expert_execution}(a), tokens from a batch may be routed to different experts based on routing decisions computed on the xPU. This is feasible due to the negligible computational cost of the routing step, which typically involves a lightweight linear layer (e.g., 4096 input and 8 output dimensions). Subsequently, only the activated experts—i.e., those assigned at least one token—are executed. Finally, the outputs from all experts are merged using a weighted sum to produce the final output tokens. Both the expert computation and result aggregation are executed by \Design~NMP processor.

The computation of a single expert in MoE models typically consists of three cascaded GeMM operations~\cite{mixtral,llama4}, as shown in Figure~\ref{fig:expert_execution}(b). Let $M$ denote the number of tokens routed to one expert in the current batch, $K$ the hidden dimension, and $N$ the intermediate dimension.  First, the input hidden matrix $\mathbf{X_1}$ of size $M \times K$ is multiplied by two weight matrices of size $K \times N$ to produce intermediate matrices $\mathbf{Z_1}$ and $\mathbf{Z_2}$ (both of size $M \times N$). A non-linear, element-wise activation is applied to $\mathbf{Z_1}$, and the result is combined with $\mathbf{Z_2}$ via a Hadamard product to form $\mathbf{X_2}$. Finally, $\mathbf{X_2}$ is multiplied by a projection-down weight matrix of size $N \times K$, producing the output $\mathbf{Z_3}$ of size $M \times K$.

\noindent \textbf{Partitioning Strategy.} In practice, different experts may receive different numbers of tokens. Furthermore, experts may be mapped to different tiers within the \dramshort~hierarchy, each with varying memory access latency, further exacerbating load imbalance. Thus, distributing multiple experts across PUs could cause serious workload imbalance issues between PUs. 
To address this, the execution of multiple chosen experts is scheduled sequentially, e.g., one expert at a time. All PUs collaborate to process one expert at a time using tensor parallelism. This requires each matrix involved in all three GeMM operations to be partitioned into tiles, each assigned to a PU for parallel execution. Figure~\ref{fig:expert_execution}(b) illustrates the matrix partitioning scheme used in \Design, where only four PUs are assumed for simplicity. Partitioning along different dimensions introduces trade-offs among input duplication, weight duplication, and partial sum aggregation. We avoid splitting along the $M$ dimension to prevent duplication of expert weights, which dominate memory usage. Instead, we split the weight matrix of the GeMM1 and GeMM2 vertically, while horizontally for GeMM3. Such a method eliminates data communication between projection-up and projection-down stages at the cost of duplicating ${\bf{X}}_t$ to multiple PUs initially and then gathering partial results from multiple PUs for ${\bf{Z}}_3$. Note that the cost of duplicating ${\bf{X}}_t$ is well amortized, as the input matrix ${\bf{X}}_1$ for all active experts is derived from ${\bf{X}}_t$ (i.e., the collection of tokens in the batch). In addition, the gathering from multiple PUs and reduction for  ${\bf{Z}}_3$ can be computed in parallel with the next expert processing, effectively hiding the latency.

\noindent \textbf{Execution Stages.} Figure~\ref{fig:expert_execution}(c) illustrates the step-by-step execution flow of the MoE layer. The xPU begins by sending the batch of input tokens, along with the corresponding expert IDs and scaling weights, to the~\dramshort and switches the \dramshort~to NMP mode (step~\callout{1}). Due to the adopted matrix partitioning strategy, each \dramshort~channel must receive the entire input token matrix. Next, the \Design~NMP processor executes the activated experts sequentially through steps~\callout{2}–\callout{7}. In steps~\callout{2} and~\callout{3}, the tensor cores in all PEs execute the two projection-up GeMM operations to compute the intermediate results $\mathbf{Z}_1$ and $\mathbf{Z}_2$. Steps~\callout{4} and~\callout{5} involve applying the activation function and performing the Hadamard product using the special function engines. Thanks to the matrix splitting strategy, no inter-PU communication is needed for each PU to obtain its required input slice for the third GeMM. The third GeMM is executed in step~\callout{6}, followed by a reduce-scatter operation to accumulate the final output matrix $\mathbf{Z}_3$ across PUs. Steps~\callout{2}–\callout{7} are then repeated for each of the remaining activated experts. In step~\callout{9}, the special function engines perform a weighted sum across expert outputs to produce the final output tokens, which are written back to the designated DRAM memory space. Finally, in step~\callout{10}, the \dramshort~exits NMP mode, and the xPU retrieves the computed tokens by accessing the designated address space.

\begin{figure}
    \centering
    \includegraphics[width=\columnwidth]{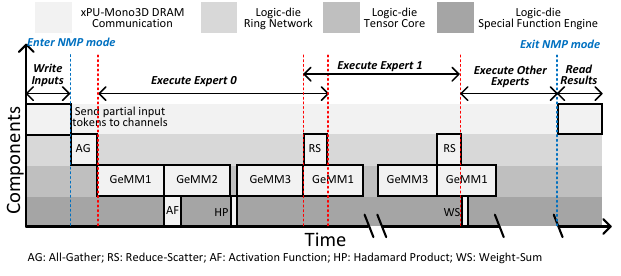}
    \caption{Optimized timing diagram of the expert processing.}
    \label{fig:expert_timing}
\end{figure}

\noindent \textbf{Execution Optimization.} Figure~\ref{fig:expert_timing} presents an optimized execution pipeline designed to maximize utilization of compute and communication resources. First, to mitigate the latency of xPU-to-\dramshort~data transfer, the input token matrix is partitioned into multiple slices, with each slice sent to a distinct \dramshort~channel. This reduces input preparation overhead, and a subsequent all-gather operation, enabled by the high-speed logic die ring network, reconstructs the full input matrix for all PUs. Second, the computation of GeMM2 is overlapped with the activation function evaluation, as there are no data dependencies between them, enabling better pipeline utilization. Third, the reduce-scatter communication associated with GeMM3 is parallelized with the GeMM1 execution of the next expert, thereby hiding communication latency behind computation. Finally, the weighted-sum operation is performed immediately by the special function engines as soon as each expert's output becomes available, minimizing idle cycles and improving overall throughput.

Within each PU, communication overhead among PEs is negligible due to the high-bandwidth shared memory. As a result, intra-PU matrix partitioning is primarily focused on maximizing tensor core mapping utilization. To this end, the longer dimension of the weight matrix is partitioned, and the resulting sub-tiles are distributed across PEs for parallel processing. Therefore, the projection-up weight slices ${\mathbf{W}}_{1,2}[i]$ are typically partitioned horizontally, while the projection-down weight slice ${\mathbf{W}}_3[i]$ is partitioned vertically across PEs to optimize compute efficiency.

\subsection{Attention Processing}\label{sec: attn_proc}

The generation task in Large Language Models (LLMs) is often bottlenecked by data access to the key–value (KV) cache. \Design addresses this issue efficiently by leveraging the high bandwidth between \dramshort and the NMP logic on the base die. However, to fully exploit this bandwidth, it is critical to effectively process the data fetched vertically from the DRAM layers on time. Otherwise, the available bandwidth may be underutilized due to computational or communication bottlenecks within the logic die.

\begin{figure}
    \centering
    \includegraphics[width=\columnwidth]{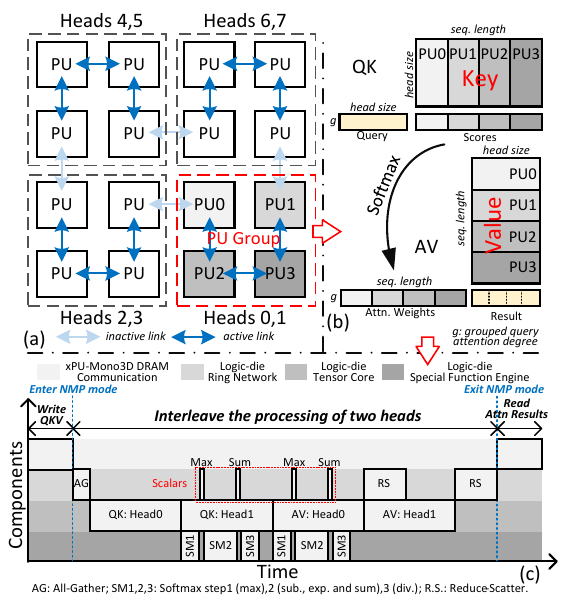}
    \caption{Execution of attention layer. (a) Heads (e.g., eight) assignment across PU groups (e.g., four). Intra-PU group: (b) Attention operator mapping. (c) Concurrent processing of multiple heads (e.g., two).}
    \label{fig:attn_PU_group}
\end{figure}

\Design leverages head-level parallelism to efficiently execute attention operations due to the absence of data dependencies across attention heads. Figure~\ref{fig:attn_PU_group}(a) illustrates the assignment of attention head tasks on the logic die. Multiple attention heads from a group of requests can be assigned across \dramshort~devices. The number of assigned heads can change depending on the network models, such as the common grouped query attention in MoE models~\cite{llama4, mixtral} and the concurrency of requests under a service latency requirement.
To provide a processing architecture for diverse head-level parallelism, the PUs on the logic die can be flexibly partitioned into multiple PU groups of variable sizes, provided that the PUs within a group are neighbors connected through the on-chip ring topology as shown in Figure~\ref{fig:attn_PU_group}(a), where PUs connected with arrows indicate the PUS on the ring. This arrangement also allows efficient intra-group communication via high-speed bi-directional links. We assign at least two heads per group to enable interleaved processing across different computation stages for the enhanced throughput and hardware utilization—for example, one head may perform a linear operation while another executes the \texttt{Softmax}. 

Figure~\ref{fig:attn_PU_group}(b) depicts how key and value matrices of a single head are partitioned across PUs within a PU group. Typically, the sequence length dimension (e.g., 512--32k tokens) is significantly larger than the attention head dimension (e.g., 64--128), motivating us to partition along the sequence length dimension. However, the \texttt{Softmax} operation inherently requires global information across all tokens, i.e., the global maximum (i.e., ${\rm row\_max}(Scores)$) and the global sum of exponentials (i.e., $\sum \exp(Scores - {\rm row\_max}(Scores))$) for normalization \cite{a3:hpca20}. Fortunately, each PU can independently compute local maxima and sums using its dedicated special function engine, requiring only scalar exchanges between PUs to derive global values. To balance the workloads of PUs in the decoding stage, the newly generated key-value pairs are distributed across different PUs within a PU group in a round-robin manner.

Figure~\ref{fig:attn_PU_group}(c) presents the optimized execution flow of multiple attention heads within a PU group. Initially, the xPU writes computed key-value pairs into the corresponding DRAM channels. Queries (which may be grouped query matrices) are partitioned into slices, each allocated to a distinct DRAM channel within a PU group.
Subsequently, all PUs in the group obtain the complete query matrix via a sub-ring all-gather operation, analogous to the MoE layer. When multiple heads are assigned to the same PU group, the \texttt{Softmax} operation can be interleaved with the $query\times key$ and $attn. \times value$ operators to minimize the overall latency. Note that the \texttt{Softmax} operator is split into three steps with two rounds of inter-PU communications as shown in Figure~\ref{fig:attn_PU_group}. Finally, the latency of the reduce-scatter of the first head can be hidden in the $attn. \times value$ operation of the second head.

In summary, \Design{} best utilizes the vertical bandwidth enabled by hybrid bonding through optimized data placement, operator mapping, and scheduling. The system applies tensor parallelism across all PU for expert computation and uses grouped-PU head parallelism for attention. Both strategies direct most memory accesses to local \dramshort{} banks through hybrid bonding I/Os.
The remaining inter-PU communication, such as all-gather, reduce-scatter, or scalar exchange, is efficiently supported by the on-chip ring network. Additionally, the scheduler overlaps matrix operations (e.g., GeMM and GeMV) with special-function computations (e.g., \texttt{SiLU} and \texttt{Softmax}), coordinating on-chip communication and compute to improve overall parallelism.

\subsection{Design with Physical Constraints}
The integration of \dramshort~and the logic die processor via hybrid bonding must satisfy both thermal and area constraints. In the NMP mode, the system could be limited by a peak power budget, $P_{peak}$, determined by thermal analysis (see \S\ref{sec:hw_eval_2}), leading to the power constraint as follows:
\begin{equation}
\begin{array}{l}
{P_{dram}} + {P_{compute}} + {P_{misc}} \le {P_{peak}},\\
{P_{dram}} = B{W_{fast\_tier}} \cdot {E_b},\;\;\,{P_{compute}} = {N_{mac}} \cdot {f_{logic}} \cdot {E_{mac}}.
\end{array}
\label{eq:power_c}
\end{equation}
Here, $BW_{fast\_tier}$ is the peak bandwidth of the fastest tier in \dramshort~tier, $E_b$ represents the energy per bit for the data transfer from the DRAM layer to the logic die via hybrid bonding, $N_{mac}$ is the total number of multiply-accumulate (MAC) units in tensor cores, $f_{logic}$ is the logic die operating frequency, and $E_{mac}$ is the energy per MAC operation. The miscellaneous power, $P_{misc}$, includes logic die SRAMs, register files, routers, special function engines, intra-PU reducers, and local memory controllers, varying according to the operator type and dataflow.

 While hybrid bonding-based data I/O does not consume an active area in the logic die, TSVs remain necessary for power delivery to both DRAM and logic dies~\cite{HB_isca24}. Consequently, the following area constraint must hold:
\begin{equation}
A_{PD} + N_{mac} \cdot A_{mac} + A_{PHY} + A_{peri} + A_{misc} \leq \alpha A_{chip}, 
\label{eq:area_c}
\end{equation}
where $A_{PD}$ is the total TSV for power delivery, $A_{mac}$ is the area per MAC unit operating at $f_{logic}$, $A_{PHY}$ represents the area of the physical communication layer of xPU-DRAM interface, $A_{peri}$ is the area of low-voltage \dramshort~peripherals on the logic die such as D/Q buffer, level shifters and others, and $A_{misc}$ captures miscellaneous logic area components similar to those outlined for $P_{misc}$, and $\alpha$ is the target utilization. Assuming a single TSV with area $A_{TSV}$ can deliver $I_{TSV}$ current, the total TSV area is given by:
\begin{equation}
\begin{array}{l}
{A_{PD}} = (\frac{{{P_{dram\_c}}}}{{{V_{dram\_c}}}} + \frac{{{P_{dram\_p}}}}{{{V_{dram\_p}}}} + \frac{{{P_{compute}} + {P_{misc}}}}{{{V_{logic}}}})\frac{{{A_{TSV}}}}{{{I_{TSV}}}},\\
{P_{dram\_c}} + {P_{dram\_p}} = {P_{dram}}
\end{array}
\label{eq:area_pd_tsv}
\end{equation}
where $V_{dram\_c}$, $V_{dram\_p}$, and $V_{logic}$ denotes the supply voltage of \dramshort~core, high-voltage peripherals, and low-voltage logic die. Equations~(\ref{eq:power_c})(\ref{eq:area_c})(\ref{eq:area_pd_tsv}) will be used to guide the design configuration of the logic die processor (see \S\ref{sec:hw_eval_3}).

\section{\Design Algorithm-System Co-Optimizations}
\label{sec:alg-sys-codesign}

\subsection{Expert Usage Prediction}
As discussed in \S\ref{sec: moe}, pre-trained MoE models often exhibit domain-specific expert specialization at inference time~\cite{yao2024exploitinginterlayerexpertaffinity}, as shown in Figure~\ref{fig:expert specialization}. Given that one of the main challenges in MoE inference is handling the large total parameter size across all experts, this specialization presents a valuable opportunity for efficient inference and serving.
When expert specialization aligns with specific query topics, it becomes possible to optimize the placement of MoE experts. For a given topic, experts with higher usage probabilities (hit rates) can be mapped to faster \dramshort tiers, reducing the latency for the data transfer from DRAM to the base logic dies.

To enable MoE expert mapping, a key component of \Design is a topic classifier that tags incoming queries. This allows the \Design scheduler to estimate the topic distribution of each query. Combined with a per-topic expert usage table (as shown in Figure~\ref{fig:serving system}), the scheduler assigns experts' weight matrices to the appropriate expert tiers. Our implementation trains a DistillBERT-based~\cite{devlin2019bert, sanh2019distilbert} topic classifier with 67M parameters on 6 topics as part of our online serving system built on \Design. 
To account for distribution shifts from standard NLP datasets to the diverse prompting styles observed in real serving queries,
we employ a data synthesis pipeline that uses GPT-4o-based rewriting to augment the training data.
Due to their compact size, our topic classifiers introduce less than 2\% latency overhead per decoding step at moderate request rates (fewer than four queries per second) on our experimental setup, while achieving 85.0\% and 81.0\% classification accuracy on real-world serving datasets (Chatbot Arena conversations~\cite{chiang2024chatbot}) for the 6-topic model, respectively. Further details on data augmentation, training, and evaluation are provided in \S\ref{sec:eval-alg}.

\subsection{Data Placement Strategy}\label{sec: data placement strategy}
\begin{figure}
    \centering
    \includegraphics[width=\columnwidth]{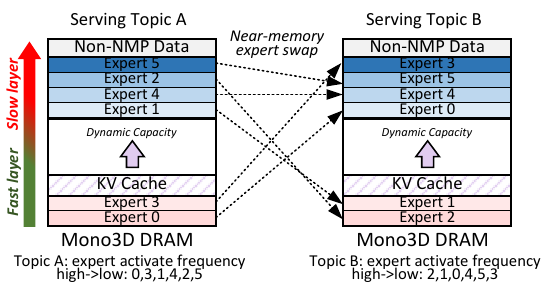}
    \caption{Example expert placement optimization for \dramshort-NMP system with tiered memory.}
    \label{fig:tier_applocation}
\end{figure}

{\setlength{\baselineskip}{0.95\baselineskip}
\begin{algorithm}[t]
\caption{\small Expert Weight Placement}
\label{alg:expert placment}
\small
\begin{algorithmic}[1]
\REQUIRE
\#Layers $L$; \#experts per layer $K$; \#active experts $k$;
usage frequencies $\mathcal{F}=\{f_{p}^{l}\mid p\!\in\![1,K],\,l\!\in\![1,L]\}$;
one expert weight size $S_E$ (bytes); DRAM banks $N_{\text{bank}}$; DRAM row-buffer size $S_{\text{rb}}$ (bytes); \#rows DRAM reserved for NMP data $\Phi$.
\ENSURE DRAM row address intervals for all expert weights $\{[a_{p}^{l}, b_{p}^{l}]\mid p\!\in\![1,K],\,l\!\in\![1,L]\}$.

\STATE $\Delta \leftarrow \left\lceil \dfrac{S_E}{N_{\text{bank}}\,S_{\text{rb}}} \right\rceil$ //\texttt{\#rows occupied by one expert}
\STATE $\tau \gets kL$ //\texttt{threshold of \#specified fast experts}
\STATE Sort $\mathcal{F}$ in descending order to obtain $\langle f_{p_1}^{l_1}, \dots, f_{p_{KL}}^{l_{KL}}\rangle$
\FOR{$i = 1$ \TO $KL$}
  \IF{$i \le \tau$}
    \STATE $a_{p_i}^{l_i} \leftarrow (i-1)\Delta$
  \ELSE
    \STATE $a_{p_i}^{l_i} \leftarrow \Phi - (KL - i + 1)\Delta$
  \ENDIF
  \STATE $b_{p_i}^{l_i} \leftarrow a_{p_i}^{l_i} + \Delta - 1$
\ENDFOR
\STATE \textbf{return} $\{[a_{p}^{l}, b_{p}^{l}]\mid p\!\in\![1,K],\,l\!\in\![1,L]\}$
\end{algorithmic}
\end{algorithm}
}

\Design~categorizes the data within the MoE model into four types: hot expert weights, cold expert weights, KV cache, and non-NMP data. Hot experts include shared experts and other experts exhibiting high routing-hit probabilities for a given topic. Non-NMP data primarily consists of miscellaneous parameters such as positional embedding parameters, layer norm shift and scale parameters, and others. These are generally used for computation in the external processor rather than the NMP. By leveraging heterogeneous access latencies across different memory tiers, a data placement strategy can be optimized to enhance the serving performance.

As shown in Figure~\ref{fig:tier_applocation}, \Design assigns non-NMP data, which is processed by the xPU, to the slowest memory tier, as accessing it requires traversing the interposer bottleneck, which is an order of magnitude slower than the internal DRAM bandwidth of the slowest tier. This helps preserve the faster memory tiers exclusively for NMP-related workloads. \Design classifies experts into hot and cold categories based on offline profiling of topic-specific requests, assigning hot experts to faster memory tiers and cold experts to slower ones. This placement ensures that hot experts benefit from low-latency access provided by faster \dramshort~memory tiers. The expert weight placement is detailed in Algorithm~\ref{alg:expert placment}. Each expert weight is partitioned into shards and distributed across \dramshort banks according to the tensor parallelism strategy (see \S\ref{sec:expert processing}). The mapping from physical row addresses obtained from Algorithm~\ref{alg:expert placment} to logical memory tiers functions as a quantization process, configurable via the tiering table (see \S\ref{sec: nmp_arch}). In our evaluation, we adopt a uniform mapping strategy that assigns an equal number of rows to each memory tier (see \S\ref{sec:hw_eval_1}). KV cache data, whose capacity dynamically changes as request generation progresses, is stored in intermediate-speed memory. Upon completing the processing of one topic (e.g., topic A), the \Design scheduler transitions to a new topic (e.g., topic B) and initiates expert swapping based on the expert activation frequencies of the new topic. To avoid costly host-processor transfers, this swapping is executed using near-memory operations, as detailed in \S\ref{sec: nmp_arch}. Specifically, the local memory controller performs the swap between two DRAM rows by temporarily buffering them in a dedicated row-swap buffer (see Figure~\ref{fig:hardware_arch}(c)) before writing them back to their new row addresses.

\section{Evaluation}
\subsection{Experimental Setup}

\subsubsection{\dramfull Configuration}\label{sec:dram_cfg}
\begin{figure}
    \centering
    \includegraphics[width=0.8\columnwidth]{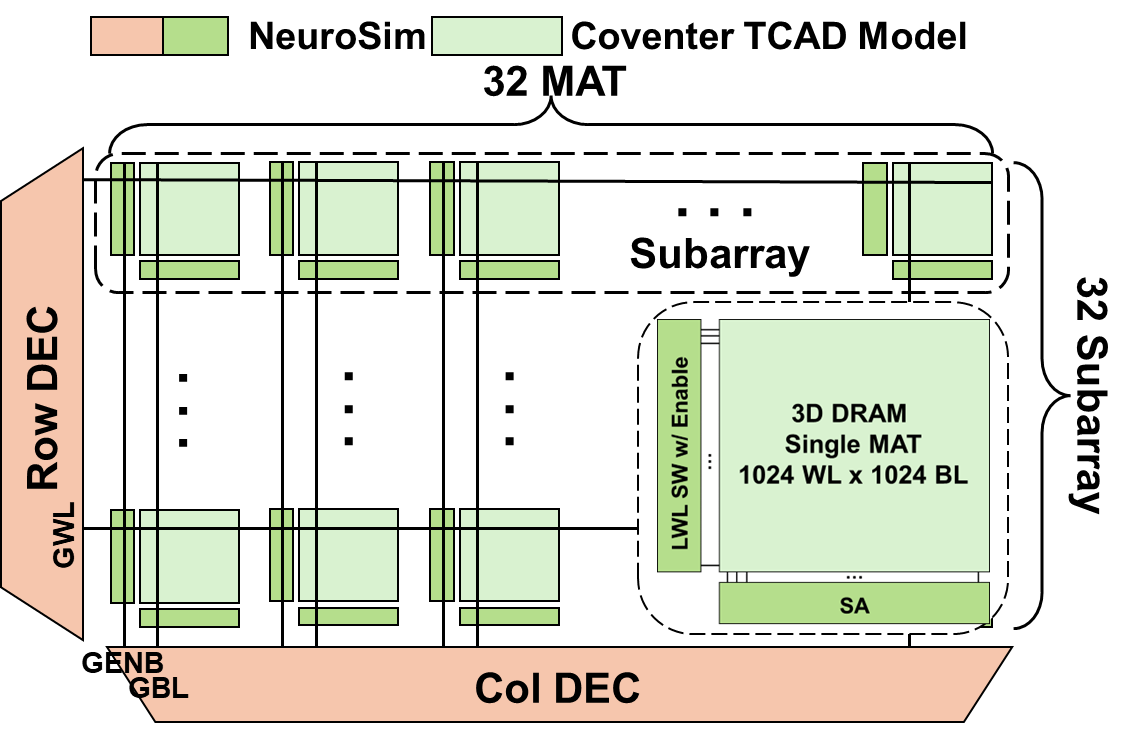}
    \caption{\dramshort bank configuration. The performance is simulated from NeuroSim \cite{Neurosim} and Coventor process simulator~\cite{coventorSemulator3D}.
    }
    \label{fig:3d dram bank}
\end{figure}

\begin{figure}
    \centering
    \includegraphics[width=0.9\columnwidth]{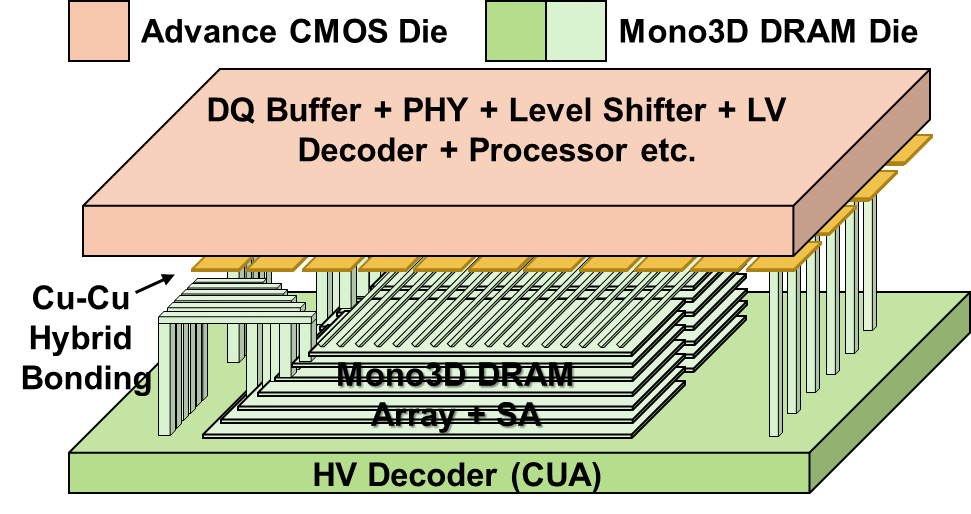}
    \caption{\dramshort array with heterogeneous integration, hybrid-bonding and CMOS-under-array (CUA).}
    \label{fig:3d dram 3d}
\end{figure}

For \dramshort technology, we adopt the vertical bitline connections for 3D stackable horizontal 1T1C. We design the \dramshort scaled to 1024 layers and define the bank structure as in Figure~\ref{fig:3d dram bank}, where 1024 BLs $\times$ 1024 WLs form a MAT and 1024 MATs form a bank. 
To illustrate the impact of heterogeneous integration, Figure~\ref{fig:3d dram 3d} presents a 3D view of the proposed \dramshort bank.
The high-voltage circuits are implemented beneath the memory array using a mature CMOS-under-array process, while the low-voltage circuits are fabricated on an advanced CMOS die and later hybrid-bonded to the memory tiers through Cu–Cu bonding pads. In this work, we leverage the 32 nm technology node for the CUA process and the 7 nm technology node for the bonded CMOS tier.
To obtain the bank-level results, we utilize the Coventor process model~\cite{coventorSemulator3D} for RC parameter extraction of the 3D DRAM array, and combine it with the peripheral circuit results extracted from NeuroSim~\cite{Neurosim} merging with the timing of DDR5 Standards \cite{jedec_ddr5}, as shown in Figure~\ref{fig:3d dram bank}. The 1T1C model of \dramshort is built by the Coventor SEMulator3D process simulator \cite{coventorSemulator3D} based on a  3D DRAM structure specification in \cite{3DDRAM_Samsung}. The detailed parameters are listed in Table~\ref{tab: 3ddram parameters}. The overall \dramshort achieves a memory density of 2.156 Gb/mm\textsuperscript{2}, which is $5.2\times$ higher than that of the latest 32Gb DDR5 die (0.417 Gb/mm\textsuperscript{2}\cite{ddr5_isscc}). It provides an internal bandwidth ranging from 19.01~TB/s to 30.34~TB/s, depending on the memory tier.

\begin{table}
    \centering
\caption{\dramfull Parameters}
\label{tab: 3ddram parameters}
\scalebox{0.84}{
\begin{tabular}{cccc}
\hline
\multicolumn{4}{c}{\textbf{\dramshort~Device Parameters}}                                                                                          \\ \hline
\multicolumn{1}{c|}{\#layers}      & \multicolumn{1}{c|}{1024}                      & \multicolumn{1}{c|}{Feature Size}    & 35 nm                       \\ \hline
\multicolumn{1}{c|}{BL/WL Pitch}   & \multicolumn{1}{c|}{70 nm/1 um}                & \multicolumn{1}{c|}{Staircase Pitch} & 500 nm                      \\ \hline
\multicolumn{1}{c|}{MAT Size}      & \multicolumn{1}{c|}{1k$\times$1k}                   & \multicolumn{1}{c|}{\#MATs/Bank}     & 32$\times$32                       \\ \hline
\multicolumn{1}{c|}{Bank Capacity} & \multicolumn{1}{c|}{1 Gb}                      & \multicolumn{1}{c|}{Bank Area}       & 0.439 mm\textsuperscript{2} \\ \hline
\multicolumn{1}{c|}{Row Buffer}    & \multicolumn{1}{c|}{32 Kb}                     & \multicolumn{1}{c|}{Energy/bit}      & 0.429 pJ                    \\ \hline
\multicolumn{1}{c|}{Chip Area}     & \multicolumn{1}{c|}{121 mm\textsuperscript{2}} & \multicolumn{1}{c|}{Chip Capacity}   & 32GB                        \\ \hline
\multicolumn{4}{c}{\textbf{\dramshort~System Parameters}}                                                                                         \\ \hline
\multicolumn{1}{c|}{Tier Design}   & \multicolumn{3}{l}{8 tiers; 4GB capacity per tier.}                                                                 \\ \hline
\multicolumn{1}{c|}{Organization} &
  \multicolumn{3}{l}{\begin{tabular}[c]{@{}l@{}}16 channels per chip (64b data I/O per channel);\\ 16 banks per channel.\end{tabular}} \\ \hline
\multicolumn{1}{c|}{DRAM Timing} &
  \multicolumn{3}{l}{\begin{tabular}[c]{@{}l@{}}tRCD={[}2.29,3.92,5.99,8.50,11.44,14.82,18.63,22.88{]} ns; \\ tRP=4.77ns; tRAS=tRCD+27.50ns; tRC=tRP+tRAS.\end{tabular}} \\ \hline

\multicolumn{1}{c|}{xPU-DRAM I/F} & \multicolumn{3}{l}{1024b data I/Os; 6.4 Gbps per pin (same as HBM3)} \\ \hline
  
\end{tabular}
}
\end{table}

\subsubsection{Logic Die Processor Modeling}\label{sec:nmp_setup}
The components of the \Design~logic die processor are implemented using SystemVerilog and synthesized using Cadence Genus~\cite{cadence_genus} with the 7nm predictive process design kit ASAP7~\cite{asap7}. The hardware employs the IEEE754 FP-16 arithmetic data format~\cite{IEEE754}, widely adopted for LLM inference serving. The local psum memory and shared memory on the logic die are implemented with SRAMs modeled by FinCACTI~\cite{fincacti}, calibrated with publicly available SRAM specifications~\cite{sram_1, sram_2}.
The area measurements for the \Design~NMP processor components are obtained from synthesis reports. Energy consumption is determined through the simulations with post-synthesis netlists, which include annotated switching activity derived from random stimulus inputs. Execution cycles, on-chip communication cycles, and associated energy metrics are derived from an in-house simulator. The simulator takes as input tensor size information, parameter tier assignments (e.g., expert parameters or KV cache), attention head mappings, and routed expert IDs, along with the delay and energy parameters for each component. It outputs the overall execution time as well as detailed energy breakdowns at the component level.

\subsubsection{System modeling}

\begin{table}
    \centering
    \caption{Evaluation Workload Setup}
    \scalebox{0.8}{
    \begin{tabular}{ccccc} \hline
         \textbf{Model}&  \textbf{Size}&  \textbf{Experts}&  \textbf{GPU Baseline}& \textbf{Stratum} \\ \hline
         OLMoE-1B-7B~\cite{olmoe}&  7B&  64 choose 8& RTX A6000& Stratum-S \\ \hline
         \mixtralsmall~\cite{mixtral}&  47B&  8 choose 2&  2$\times$H100& Stratum-L\\ \hline
         Qwen2.5-32B~\cite{yang2024qwen2}&  32B&  Non-MoE&  2$\times$H100&  Stratum-L \\ \hline
 Llama-4-Scout~\cite{llama4}& 109B& \makecell{1 shared \\ + 16 choose 1}& 4$\times$H100& Stratum-XL\\ \hline
    \end{tabular}
    }
    \label{tab:eval models}
\end{table}

We evaluate with models (both MoE and regular LLMs) and system configurations shown in Table~\ref{tab:eval models}. Each GPU baseline and \Design configuration is chosen to support the maximum evaluated context length without degrading performance. The GPU baselines are evaluated using vLLM 0.8.1 \cite{vllm} under benchmark throughput mode using NVIDIA RTX A6000 or H100 SXM5 HBM3 GPUs for different \Design configurations. The GPU energy is derived from the NVIDIA-SMI tool.

The system-level simulator contains a Request Generator, SLO-Aware Scheduler, Memory and Computation Mapper, and interfaces to \Design NMP simulator, in accordance with Figure~\ref{fig:serving system}. The Request Generator models a Poisson process in which the incoming queries of certain topics arrive at defined rates. Taking into consideration serving SLO, the scheduler dynamically batches input queries to the \Design processor for inference and prioritizes dispatching input queries of the same topic to maximize hot expert hits. Using the prior knowledge of the expert usage table, the memory mapper aggregates the topics in the batch and calculates expert placements for \dramshort that maximize hot expert hit, as shown in Algorithm~\ref{alg:expert placment}. A memory reconfiguration is executed between dispatches to relocate experts. Energy and latency consumed by xPU and NMP are accumulated during simulated serving. 

\subsection{Hardware Evaluation}\label{sec:hw_eval}
\subsubsection{Tiering in 3D-DRAM}\label{sec:hw_eval_1}
\begin{figure}
    \centering
    \includegraphics[width=\columnwidth]{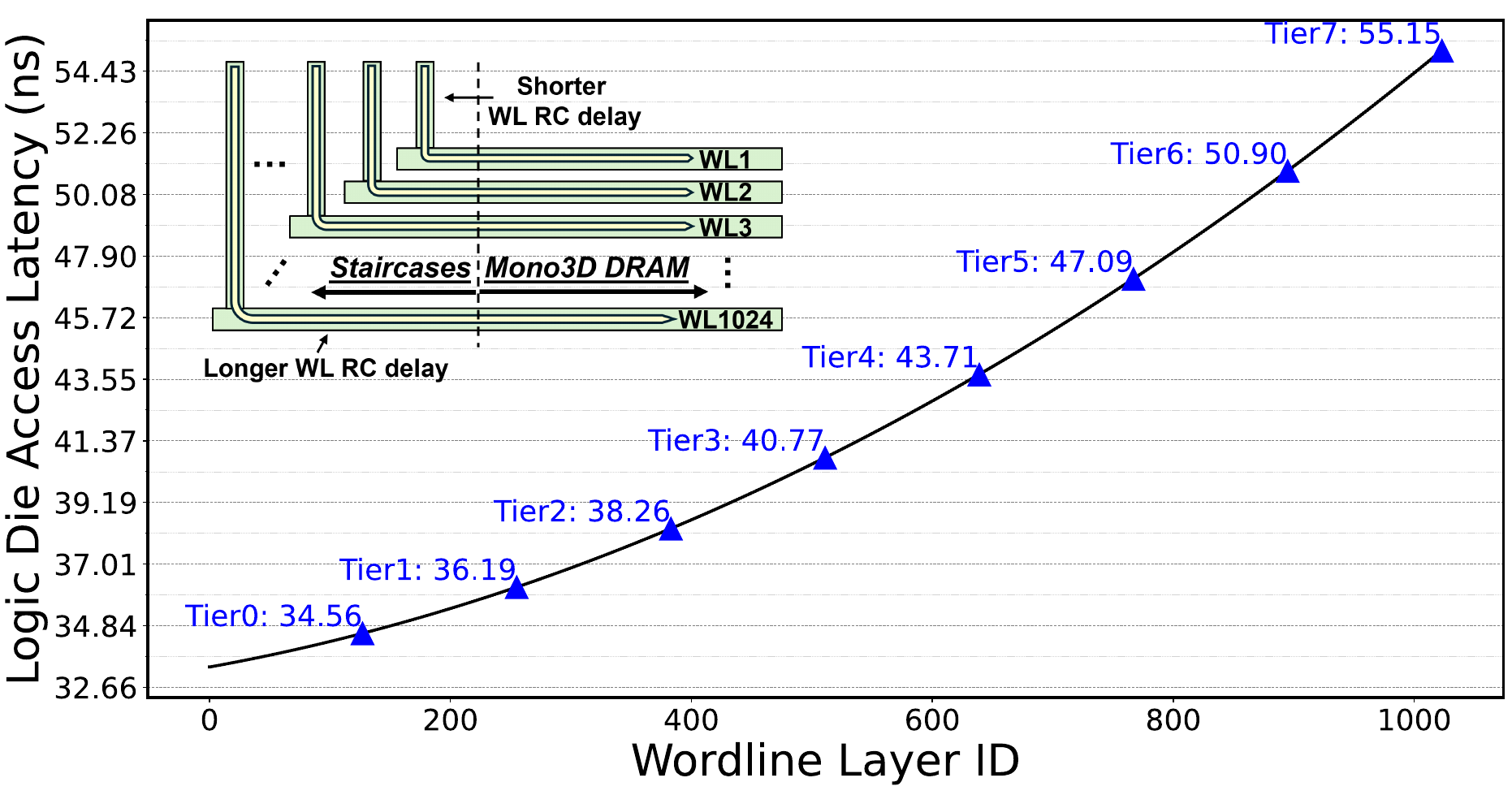}
    \caption{\dramshort latency across WL layers. The inset illustrates various access latencies according to the increasing WL RC delay when scaling the staircase for increasing WL layers. 
    }
    \label{fig:3d dram latency}
\end{figure}
As illustrated in Figure~\ref{fig:3d dram latency}, \dramshort exhibits the almost linearly scaled access latency associated with the extending WL staircase structure for accessing various WL layers. As \dramshort vertically scaled with increasing WL layers, WL parasitics corresponding to the area of the staircase are also scaled, leading to a longer RC delay. Although the critical path for the bottommost WL suffers from long latency, the topmost WL has a shorter access latency, facilitating further optimization at the system level. In this work, we introduce the memory tiering technique for \dramshort. We define 8 timing tiers in \dramshort corresponding to different layers as shown in Figure~\ref{fig:3d dram latency}. The fast tier achieves 1.6$\times$ faster access than the slowest tier. 

\subsubsection{Power and area budget.}\label{sec:hw_eval_2}

\noindent\textbf{Power.} The vertically integrated memory and logic dies require precise thermal modeling to determine the logic die's power budget. We performed thermal simulations using the HotSpot~\cite{hotspot_isca03, hotspot_te03} simulator for 3D IC. We consider high-end liquid cooling solutions with vapor chamber heat sinks. The heat sink is characterized by the following parameters: a convection capacitance of 75~J/K, a convection resistance of 0.01~W/K, and a thickness of 1~mm. The material properties include a thermal conductivity of 5000~J/(m$\cdot$K) and a specific heat capacity of $10^6$~J/(m$^3\cdot$K). The thermal conductivity values are adopted from previous studies on vapor chamber thermal modeling~\cite{vchs_5000wmk, celsia_heatsink}. Additionally, advanced cooling fluids, such as phase change materials, achieve significantly reduced convection resistance of approximately $\mathrm{0.01 \,W/K}$~\cite{vchs_001wk_survey, vchw_001wk_2}. Furthermore, we derived convection capacitance, heat sink thickness, and vapor specific heat parameters, explicitly considering the differences between conventional and vapor chamber heat sinks. Prior research demonstrates that state-of-the-art cooling methods for 3D ICs effectively manage power densities ranging up to $\mathrm{200\,W/cm^2}$~\cite{kim2010thermal}. Assuming full utilization of \dramshort~internal bandwidth at 30.34 TB/s, each \dramshort~die consumes approximately 104~W. Given the safe temperature for memory and data~\cite{han2021power_dram85C}, we conclude the logic die power caps at around 45W per chip.

\noindent\textbf{Area.} The \dramshort~maintains compatibility with the xPU-DRAM interposer interface utilized by HBM3~\cite{hbm3}, thereby requiring an HBM3 PHY module. The PHY module's area overhead, computed for 16 physical channels each supporting 64-bit data I/O at 6.4 Gbps, totals 23.94 mm\textsuperscript{2}~\cite{HBM3_phy, vlsi_scaling}. The logic die also has low-voltage \dramshort~peripherals such as DQ buffer, level shifter, and address decoder, occupying 14.80 mm\textsuperscript{2}. Power delivery to both \dramshort~and the logic dies involves TSVs extending through the logic die from the interposer. Each TSV with an area of 25~$\mu$m\textsuperscript{2} can deliver up to 36 mA~\cite{HB_isca24}. To accommodate peak power of 104~W for the \dramshort~and 45W for the logic processor, the TSVs introduce an area overhead of 0.21 mm\textsuperscript{2} when considering a 2:1 redundancy scheme. The logic die matches the \dramshort~die area of 121 mm\textsuperscript{2} (i.e., the base die dimensions of HBM3~\cite{hbm3}). Thus, the available area budget for the logic die processor is 82~mm\textsuperscript{2}.

\subsubsection{logic die processor.}\label{sec:hw_eval_3}
\begin{table}[t]
\centering
\caption{\Design Logic Die Processor Specification}
\label{tab:spec_nmp}
\scalebox{0.75}{
\begin{tabular}{cccc}
\hline
\multicolumn{4}{c}{\textbf{Processing Element (PE)}}                                                                                             \\ \hline
\multicolumn{1}{c|}{Tensor Core}          & \multicolumn{1}{c|}{16$\times$16MACs}  & \multicolumn{1}{c|}{Tiering Table}   & 16$\times$16b Registers \\ \hline
\multicolumn{1}{c|}{Psum SRAM}            & \multicolumn{1}{c|}{64 KB}        & \multicolumn{1}{c|}{Row Swap Buffer} & 8KB RF           \\ \hline
\multicolumn{4}{c}{\textbf{Processing Unit (PU)}}                                                                                                \\ \hline
\multicolumn{1}{c|}{\#PEs}                & \multicolumn{1}{c|}{16}           & \multicolumn{1}{c|}{Shared Memory}   & 1.25 MB          \\ \hline
\multicolumn{1}{c|}{Special Func. Engine} & \multicolumn{1}{c|}{256-way SIMD} & \multicolumn{1}{c|}{Ring Router}     & 128 GB/s/link       \\ \hline
\multicolumn{4}{c}{\textbf{\Design NMP Processor}}                                                                                \\ \hline
\multicolumn{1}{c|}{Basic}                & \multicolumn{3}{c}{7 nm process; 0.7 V supply; 121 mm\textsuperscript{2} die area; FP16 format.}                           \\ \hline
\multicolumn{1}{c|}{\#PUs}                & \multicolumn{1}{c|}{16}           & \multicolumn{1}{c|}{SRAM Capacity}   & 36 MB            \\ \hline
\multicolumn{1}{c|}{Peak Performance}     & \multicolumn{1}{c|}{128 TFLOPS}   & \multicolumn{1}{c|}{Peak Power}      & 43 W             \\ \hline
\multicolumn{1}{c|}{\begin{tabular}[c]{@{}c@{}}Aggregated On-chip\\ Ring Bandwidth\end{tabular}} &
  \multicolumn{1}{c|}{2.048 TB/s} &
  \multicolumn{1}{c|}{\begin{tabular}[c]{@{}c@{}}Aggregated Mono3D\\ DRAM Bandwidth\end{tabular}} &
  19.01-34.34 TB/s \\ \hline
\end{tabular}
}
\end{table}

\begin{figure}
    \centering
    \includegraphics[width=0.9\columnwidth]{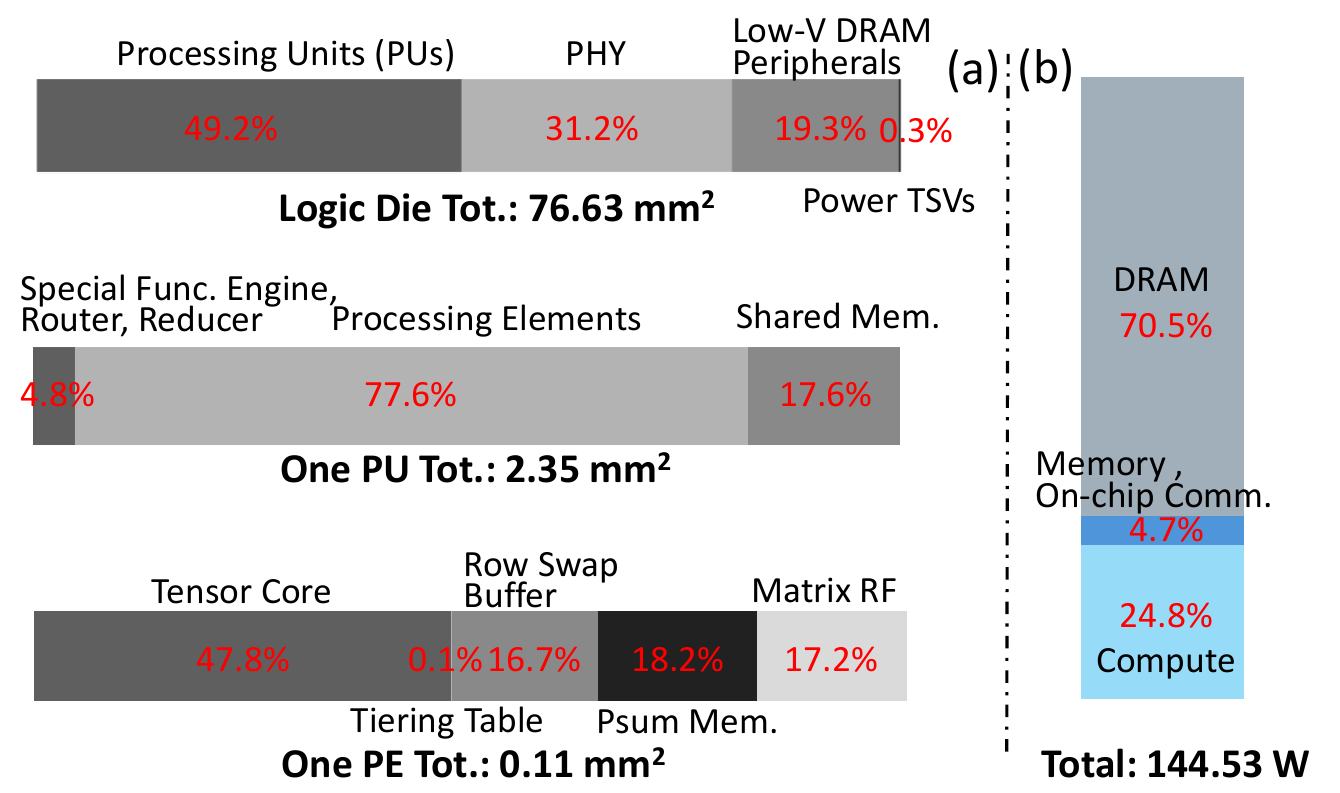}
    \caption{(a) Area breakdown of logic die processor; (b) Power breakdown of \dramshort-Logic Die at peak performance.}
    \label{fig:nmp_breakdown}
\end{figure}
\noindent Table~\ref{tab:spec_nmp} summarizes the specifications of the \Design~logic die processor at the PE, PU, and chip hierarchy levels. We calculated the maximum number of MAC units using Equation~(\ref{eq:power_c}), employing a simulated per-MAC-operation energy of $E_{mac}=0.604$ pJ. The processor achieves a peak performance of 128 TFLOPS with 64k MAC units operating at 1 GHz. The PE tensor core is arranged into a 16$\times$16 array, providing a balanced matrix tile size to optimize utilization across diverse GeMM sizes. Additionally, a programmable tiering table stores row addresses of the last \dramshort~layer and the tRCD for each tier. The incoming row addresses are compared with eight stored addresses to expedite tRCD lookup. The communication-computation optimizations adopted enable the on-chip ring to require only 128 GB/s bandwidth per link without performance degradation based on the system-level simulation. Figure~\ref{fig:nmp_breakdown} presents the area and power breakdown of the \Design~NMP stack. The total area occupied by the active logic is 76.63 mm\textsuperscript{2}, which falls within the 121 mm\textsuperscript{2} area budget, yielding a utilization of 63\%. The area is predominantly consumed by the PEs, which dominate the PU-level area. The tiering table introduces only a minimal overhead of 0.1\% of the PE area within each PE. The \Design~NMP stack reaches a peak power of 144.53 W when the fastest \dramshort~tier is accessed concurrently with full tensor core utilization. The total power of the logic die is 42.67~W, including compute, on-chip communication, and logic-die memory access, under the 45W power budget.

\subsection{System Evaluation}

\subsubsection{Algorithm Evaluation}
\label{sec:eval-alg}

\begin{figure*}
    \centering
    \includegraphics[width=2.1\columnwidth]{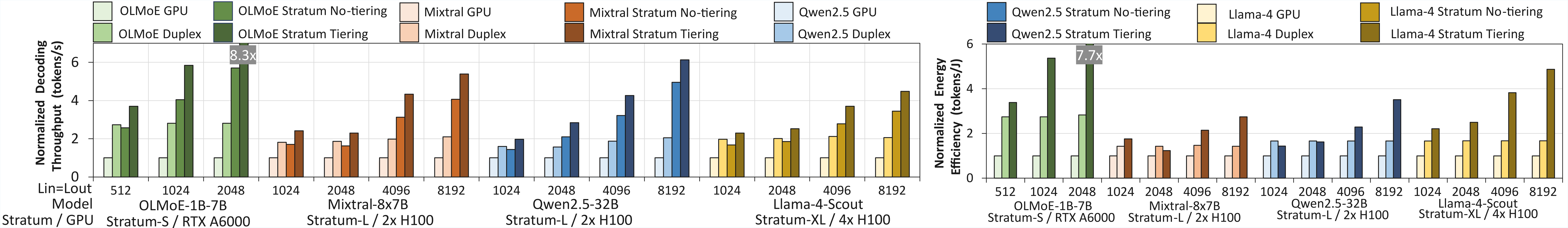}
    \caption{Evaluation and comparison of system decoding throughput and energy efficiency.}
    \label{fig: overall tpt}
\end{figure*}

\textbf{Model.} Our model is based on DistilBERT~\cite{sanh2019distilbert} with 67M parameters and designed for multi-topic text classification, supporting sequences of up to 1024 tokens. It features a compact architecture with 6 transformer layers and 12 attention heads, with a hidden dimension of 3072. 

\noindent \textbf{Data.} Our model training involves a customized data mix across 6 topics. The datasets include a 2\% split of Pile of Law for legal topic~\citep{henderson2022pile}, 1 out of 3 splits from atlas converse and INCLUDE for humanity topic~\cite{atlas_converse, romanou2024include}, 5\% split of Programming books for CS topic~\citep{openphi_programming_books_llama_2025}, SciQ and ARC-easy for science topic~\citep{2017sciq, Clark2018ThinkYH}, GSM8K and MATH for math topic~\citep{cobbe2021gsm8k, hendrycksmath2021}, Atlas reasoning for logic topic~\citep{atlas_reasoning_2025}. For the above-mentioned 6-topic configuration, the data encompasses approximately 70 million tokens. 

\noindent  \textbf{Training and Evaluation.} To address distribution shifts from standard NLP datasets to diverse real-world prompts, we use a GPT-4o-based data synthesis pipeline. We sample 500 prompts from the Chatbot Arena dataset~\cite{chiang2024chatbot} to reflect natural user styles, then use GPT-4o with a fixed system prompt to rewrite 50\% of our training data into a QA format. We use a mix of rewritten and original data to train our topic classifier on a single A100 GPU for 3 epochs of 3 hours each. For evaluation, we use the MMLU test sets~\citep{hendryckstest2021} and hand-curated 180-example subsets of Chatbot arena conversations dataset~\citep{chiang2024chatbot} with the 6 topics. Our trained classifier achieves 94.5\% and 85.0\% accuracy on MMLU and Chatbot arena test sets, close to the performance of OpenAI O3-mini-high (96.2\%, 91.1\%). The inference overhead of the model is less than 10ms with ONNX runtime on a regular laptop CPU. 
We use OpenAI-O3 LLM-as-a-judge to classify 33,000 real-world queries from LMArena~\cite{lmarena}, which shows that our six coarse-grained topics cover 93\% of queries, confirming the robustness and generality of TopicBERT’s taxonomy.

\subsubsection{System Performance}\label{sec:eval system} 
Figure~\ref{fig: overall tpt} shows the normalized decoding throughput and energy efficiency when serving requests with equal input and output length. For \dramshort designs, we evaluate \textit{no-tiering} and \textit{tiering} approaches. In \textit{no-tiering} design of \dramshort, \dramshort is treated as a single tier, therefore, the logic die is limited to operating under the worst memory access latency of the memory die. In \textit{tiering}, \dramshort is divided into 8 tiers with fine-grained memory latency and data mapping optimizations given tiering. \Design \textit{tiering} consistently outperforms GPU baselines across all cases, averaging 8.29$\times$, 5.39$\times$, 6.13$\times$, 4.48$\times$ better decoding throughput for OLMoE, Mixtral, Qwen2.5, and Llama-4, respectively. Specifically, as decoding length grows, decoding on conventional GPUs with limited memory bandwidth becomes increasingly memory-bound, due to the quadratic complexity of the attention mechanism, explaining the growing gap of \Design over GPU baselines. \Design \textit{no-tiering} as well outperforms GPU due to its higher internal bandwidth compared to HBM, even considering the worst-case latency. The internal memory tiering (\S\ref{sec: nmp_arch}) and MoE-specific data mapping optimizations (\S\ref{sec: data placement strategy}) further improve decoding throughput by averages of 1.45$\times$, 1.39$\times$, 1.32$\times$, 1.34$\times$ over \textit{no-tiering} for the 4 models, respectively. Energy-wise, \Design achieves up to 7.66$\times$, 2.74$\times$, 3.51$\times$, 4.87$\times$ better energy efficiency for the same decoding tasks across OLMoE, Mixtral, Qwen2.5, and Llama-4, respectively, due to cheaper memory access. We also extracted data from the previous work Duplex~\cite{duplex_micro24} and made conservative 
scaling to compare with \Design. \Design achieves up to 2.9$\times$, 2.5$\times$, 3.0$\times$, 2.2$\times$ better throughput and 2.7$\times$, 1.9$\times$, 2.9$\times$, 2.1$\times$ energy over Duplex~\cite{duplex_micro24} for OLMoE, Mixtral, Qwen2.5, and Llama-4.

\subsubsection{Expert Placement Optimizations}
\label{sec: expert placement}

\begin{figure}
    \centering
    \includegraphics[width=\columnwidth]{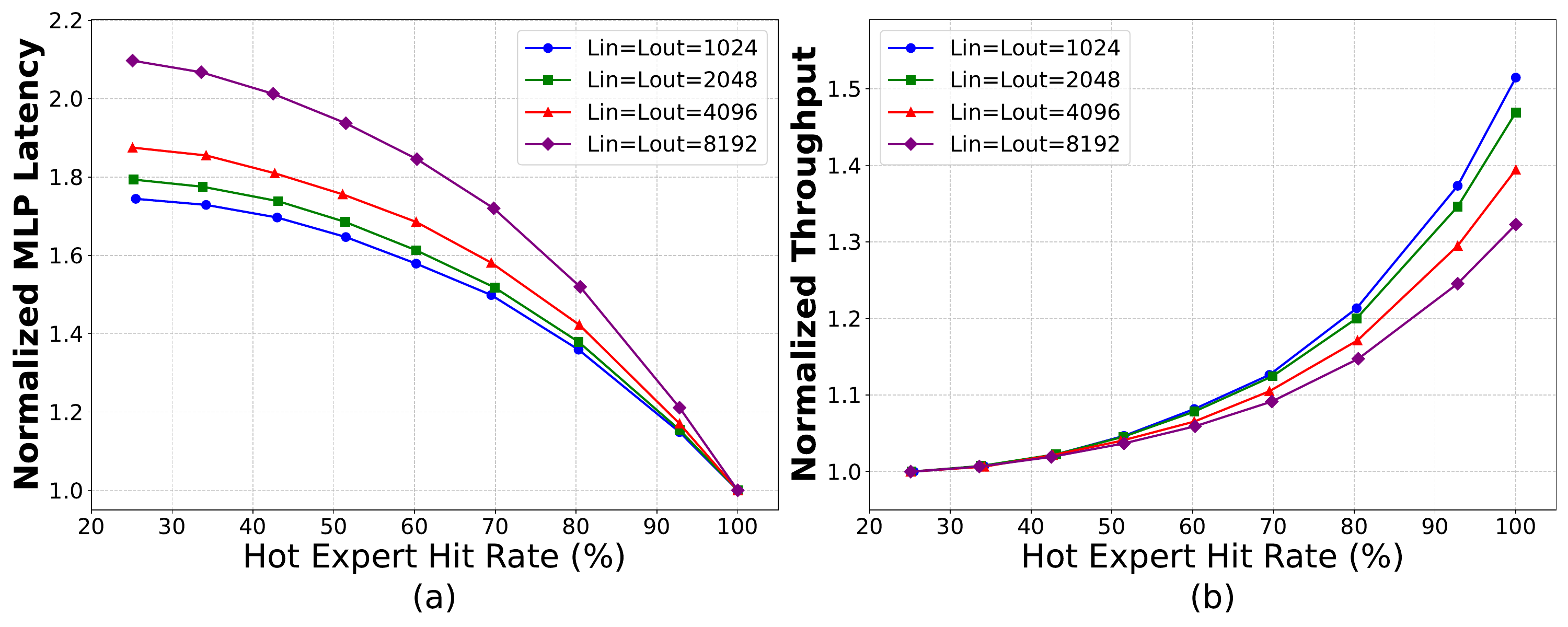}
    \caption{Impact of hot expert hit rates on (a) MLP (MoE layer) latency and (b) overall system throughput for \Design-L.}
    \label{fig: hit rate vs tput}
\end{figure}

\textbf{Effectiveness.}
   To study the effectiveness of expert placement in the tiered \dramshort, we scan the hot expert hit rate for \mixtralsmall on \Design-L as shown in Figure~\ref{fig: hit rate vs tput}. The hot expert hit rate is defined as the ratio of aggregated hot expert to total expert accesses at the token level. Across decoding lengths, accurate hot expert usage prediction brings 1.32$\times$ to 1.51$\times$ better throughput over a uniformly distributed expert usage, or equivalently a naively managed tiered memory. The benefit is more noticeable on smaller decoding lengths, as the MLP dominates the decoding latency more. Using our topic prediction model, we achieve 31.6\%, 48.5\%, and 68.9\% aggregated hot expert hit rates when serving Mixtral, OLMoE, and Llama-4.

\begin{table}[t]
\centering
\caption{Overhead of Expert Swap across \dramshort Tiers}
\label{tab:swap_cost}
\scalebox{0.8}{
\begin{tabular}{c|c|c|c}
\hline
          & \textbf{OLMoE}~\cite{olmoe} & \textbf{Mixtral}~\cite{mixtral} & \textbf{Llama-4}~\cite{llama4} \\ \hline
\#Expert swaps/sec & 5.91        & 2.59         & 4.02          \\ \hline
Time Overhead (ms)   & 0.64 (0.37\%)            & 0.90 (0.23\%)          & 0.45 (0.18\%)          \\ \hline
Energy Overhead (mJ) & 0.25 (\textless{}0.02\%) & 0.35 (\textless{}0.03\textperthousand) & 0.34 (\textless{}0.02\textperthousand) \\ \hline
\end{tabular}
}
\end{table}

\noindent \textbf{Costs.}
The scheduler (\S\ref{sec: serving}) may trigger expert swaps \textit{between batches}. To evaluate the worst-case scenario, we consider 1) short sequences, ${L_{in}}={L_{out}}=256$ with batch size one, and 2) consecutive batches assigned to different topics. Table~\ref{tab:swap_cost} reports the time and energy overheads of expert swaps, which remain well below 1\% across all benchmarks. 
This negligible cost stems from two factors: expert swaps occur within the same bank, avoiding cross-bank movement, and NMP logic includes dedicated row-swap buffers that enables swapping at the high internal \dramshort tier bandwidth without traversing the DRAM–xPU interface.

\subsubsection{Performance scaling with batch size}
\label{sec: ablation batch size}

\begin{figure}
    \centering
    \includegraphics[width=\columnwidth]{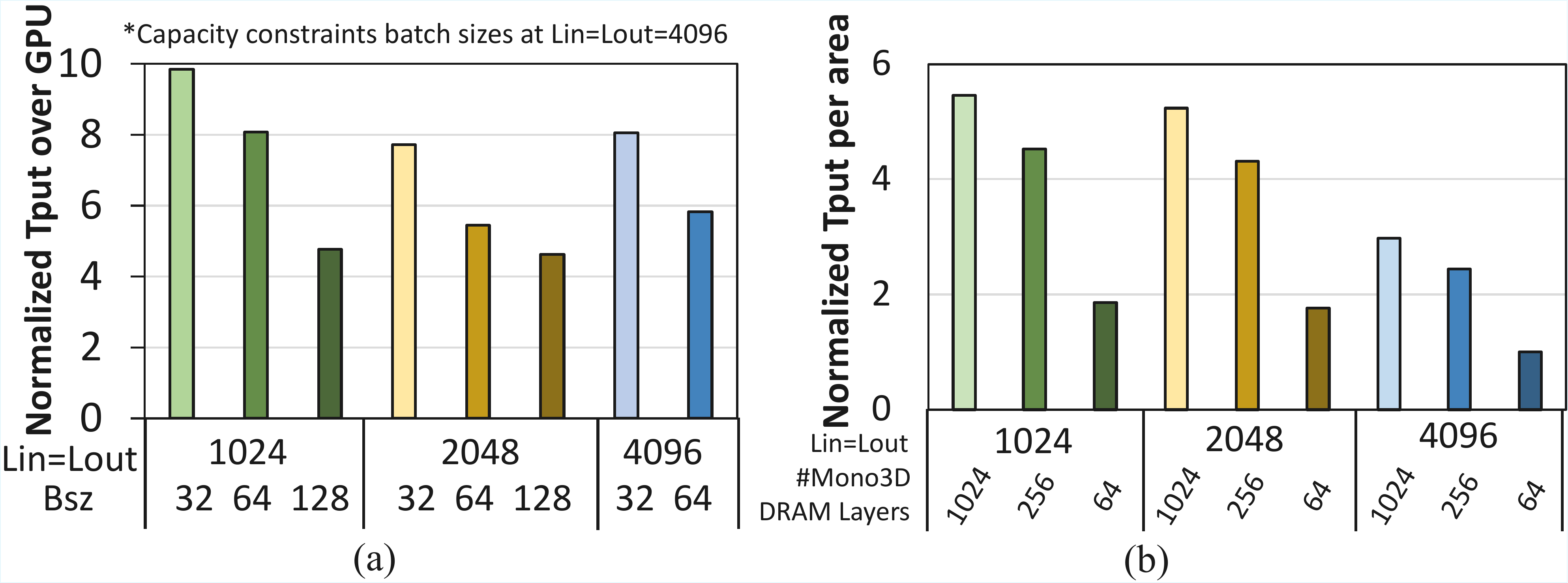}
    \caption{Impacts of (a) batch size and (b) \dramshort layers on system-level metrics, evaluated with Llama-4-Scout on \Design-XL}.
    \label{fig: ablation batch and layers}
\end{figure}

Figure~\ref{fig: ablation batch and layers}(a) evaluates \Design's performance scaling across different query batch sizes using the large-scale Llama-4-Scout~\cite{llama4} benchmark. Batch sizes are chosen to ensure the full model fits within the \dramshort of \Design or the HBM of the GPU baseline. \Design consistently outperforms the GPU baseline across all settings by 4.7--9.8$\times$. However, the relative performance advantage reduces with larger batches, particularly at shorter sequence lengths (e.g., 1024 tokens), due to the GPU die’s higher compute-to-bandwidth ratio and the increased dominance of MoE layers in the overall runtime. 
\subsubsection{Performance scaling with \dramshort layers}
\label{sec: ablation layers}
Figure~\ref{fig: ablation batch and layers}(b) reports \Design's performance scaling across different \dramshort layer configurations. All variants have the same DRAM capacity and use the same NMP logic die processor, and throughput is normalized to the die area of each \dramshort to ensure a fair, cost-aware comparison. On average, the 1024-layer design achieves 1.21$\times$ and 2.96$\times$ higher throughput per area than the 256-layer and 64-layer \dramshort, respectively, demonstrating the cost-efficiency benefits of adopting >1k-layer \dramshort.

\subsubsection{Tiering mechanism on \dramshort with less layers.}\label{sec: tiering at lower layers}
The proposed tiering mechanism exploits wordline latency variation resulting from vertical stacking in monolithic 3D DRAM. \dramshort employs the similar fabrication process as 3D NAND Flash, which has already scaled beyond 400 layers~\cite{Samsung_3DNAND_400}. Thus, we consider a 512-layer configuration by partitioning the original 1024-layer mat into two horizontally connected 512-layer segments while preserving the NMP logic design. Device-level simulations reveal a 1.3$\times$ access latency difference between the fastest and slowest tiers. System-level evaluations demonstrate overall (including both MoE and attention layers) performance improvements of 17.7\%, 18.3\%, and 18.3\% under our topic-aware tiering placement at a sequence length of ${L_{in}} = {L_{out}} = 1024$ on LLama-4-Scout~\cite{llama4}, Mixtral 8$\times$7B~\cite{mixtral}, and OLMoE-1B-7B~\cite{olmoe} benchmarks, respectively. These results validate the efficacy of the proposed tiering strategy across a wide number of \dramshort layers.

\section{Related Works}
\noindent \textbf{3D Stackable DRAM.}
\dramfull has emerged as a promising alternative to HBM by sequentially fabricating multiple DRAM layers on the same wafer. 
Unlike HBM, which depends on TSVs and costly die-stacking, \dramshort\ employs fine-pitch hybrid bonding for higher internal bandwidth and integration density~\cite{3DDRAM_Samsung, huang20233d, chen20223d, chen2023highly, 3DDRAM_Stanford, 3DDRAM_GT}.
Leading \dramshort\ technologies include Horizontal 1T1C~\cite{3DDRAM_Samsung, huang20233d}, which reorients and stacks 1T1C DRAM cells, and Gate-Control Thyristors~\cite{chen20223d, chen2023highly}, which leverage avalanche mechanisms. 
Recent work further shows that \dramshort’s $\sim$1$\mu$m bonding pitch~\cite{CuCuHyBond} enables up to 5$\times$ denser vertical interconnects than HBM~\cite{HB_isca24}.

\noindent \textbf{Processing In/Near Memory Acceleration for Transformers.}
While Processing In/Near Memory (PIM/PNM) has been a long-standing concept, MAT~\cite{zhou2021mat} first applied PIM to Transformer models, targeting a single encoder block with a memory-efficient pipelined sub-sequence flow.
TransPIM~\cite{zhou2022transpim} extends this with a hybrid PIM-PNM architecture for full-model execution. Neupims~\cite{heo2024neupims} and AttAcc~\cite{attacc_asplos24} focus on Decoder-only Transformer models, offloading attention layers in the decoding stage to the PNM on a xPU-PNM hybrid-processing system. Duplex~\cite{duplex_micro24} further expanded support to MoE, GQA, and continuous batching with dynamic compute partitioning. However, all these designs rely on 2D DRAM or die-stacked HBM, limiting their effectiveness when applied to \dramshort-based systems.

\section{Conclusion}

We present \Design, a novel system–hardware co-design for efficient MoE serving that, for the first time, leverages high-density \dramshort{} dies integrated with logic through 3D hybrid bonding, and further connected to GPUs via a 2.5D silicon interposer. This architecture offers a cost-effective and high-throughput alternative to conventional GPU–HBM-based systems. At the hardware level, \Design introduces in-memory tiering to exploit vertical access latency variations in \dramshort{}, and a near-memory processor (NMP) optimized for expert and attention execution. At the system level, we exploit topic-dependent expert activation patterns to classify and map experts across memory tiers and design a topic-aware scheduler guided by a lightweight classifier to meet service-level objectives. Cross-layer evaluations spanning device, circuit, algorithm, and system levels show that \Design achieves up to 8.29$\times$ better decoding throughput and up to 7.66$\times$ less energy consumption compared to GPU baselines.

\begin{acks}  
This work was supported in part by PRISM and CoCoSys, centers in JUMP 2.0, an SRC program sponsored by DARPA. This research is also  supported by National Science Foundation (NSF) grants 2112665,  2112167, 2003279, 2120019, and 2211386.
\end{acks}

\newpage

\bibliographystyle{ACM-Reference-Format}
\bibliography{reference}

\end{document}